\documentclass[aps,prd,twocolumn,showpacs,10pt,superscriptaddress,preprintnumbers,nofootinbib]{revtex4-1}
\usepackage{epsfig,amssymb,amsmath,psfrag,epstopdf,color}
\pdfoutput=1
\usepackage{graphicx}
\usepackage{multirow}
\usepackage{subcaption}
\usepackage{ragged2e}
\usepackage[normalem]{ulem}
\usepackage[dvipsnames]{xcolor}
\DeclareCaptionJustification{justified}{\justifying}
 \usepackage[normalem]{ulem}
\usepackage{paralist}
\usepackage{hyperref}
\usepackage[size=tiny]{todonotes}

\usepackage{float}
\usepackage[multidot]{grffile}

\allowdisplaybreaks

\def\eqref#1{{Eq.\!~(\ref{#1})}}
\def\figref#1{{Fig.\!~\ref{#1}}}
\def\secref#1{{Sec.\!~\ref{#1}}}

\newcommand\tauf[1][]{$\tau_\text{form}$}
\newcommand\tauff[1][]{\tau_\text{form}}
\newcommand\xjz[1][]{$x_{j,Z}$}
\newcommand\dR[1][]{$\Delta R$}
\newcommand\dRR[1][]{\Delta R}
\newcommand\mg[1][]{$m_{g}$}

\begin{document}

\title{Exploring the time axis within medium-modified jets}

\author{Liliana Apolinario}\email{liliana@lip.pt}
\affiliation{LIP, Av.\ Prof.\ Gama Pinto, 2, P-1649-003 Lisbon, Portugal}
\affiliation{Instituto Superior T\'{e}cnico (IST), Universidade de Lisboa, Av.\ Rovisco Pais 1, 1049-001, Lisbon, Portugal}

\author{Pablo Guerrero-Rodr\'iguez}\email{pguerrero@lip.pt}
\affiliation{LIP, Av.\ Prof.\ Gama Pinto, 2, P-1649-003 Lisbon, Portugal}

\author{Korinna Zapp}\email{korinna.zapp@fysik.lu.se}
\affiliation{Department of Physics, Lund University, Box 118, SE-221 00 Lund, Sweden}

\date{\today}

\begin{abstract}
In this manuscript, we illustrate how to use the newly proposed $\tau$ re-clustering algorithm to select jets with different degrees of quenching without biasing their initial transverse momentum spectrum. Our study is based on Z+jet simulated events using the JEWEL Monte Carlo event generator to account for jet quenching effects. We apply the $\tau$ re-clustering algorithm to extract a proxy for a time axis (formation time) within the evolving medium.
This information allows us to label jets according to their fragmentation pattern and select populations with enhanced sensitivity to quenching effects. Our results illustrate the potential of jets as precision tools for QGP tomography. Further, we show that the discussed method minimizes the biases stemming from $p_{T}$-, \dR- or mass-based jet selection.
\end{abstract}

\let\clearpage\relax
\maketitle

\section{Introduction}

The physics program of current heavy-ion colliders is focused on determining the characteristics of the hot and dense matter that is formed in these systems: the Quark-Gluon Plasma (QGP). Among the different experimental signatures of QGP formation, \emph{jet quenching}, the collection of energy loss processes~\cite{CMS:2011iwn,ATLAS:2010isq,CMS:2021vui,ALICE:2018vuu,ATLAS:2018gwx,STAR:2023pal,STAR:2020xiv} and modification of the jet substructure~\cite{ATLAS:2022vii,ALICE:2021mqf,STAR:2021kjt,CMS:2017qlm,CMS:2021otx,CMS:2013lhm,ATLAS:2018bvp,CMS:2018fof,ALICE:2021mqf} resulting from the jet-medium interactions, offers unique insights into QGP formation~\cite{Apolinario:2022vzg}. Jets evolve simultaneously with the QGP and thus offer a unique capability to explore distinct phases of the collision, thereby potentially contributing to a better understanding of QGP onset conditions.

Even though the fast evolution of the QGP makes its interaction with jets an inherently time-dependent process, this crucial dimension is missing from current jet quenching measurements, limiting the assessment of the produced medium to an average quantification of its properties. This has motivated considerable theoretical interest in the use of jet quenching to study the evolution of local medium properties~\cite{Apolinario:2017sob,Attems:2022ubu,Wang:2002ri,He:2020iow,Andres:2022bql,Barata:2023qds}. Most recently, jet substructure observables have been proposed as a tool to access the QGP time structure. In a recent work \cite{Apolinario:2020uvt} we showed that by using a novel jet clustering algorithm (the $\tau$ algorithm) it is possible to define an adequate proxy for the formation time, \tauf, to enhance energy loss effects. This allows to classify jets according to their sensitivity to medium interactions, an application explored in Ref.\ \cite{Apolinario:2020uvt} through the calculation of nuclear modification factors in Monte Carlo-simulated dijet events. In the present work we delve deeper into the $\tau$ algorithm, and on $\tau_{\text{form}}$ as jet substructure observable, unveiling some unique features that strengthen its value as a promising tool for QGP tomographic measurements. We do this by applying the techniques presented in Ref.\ \cite{Apolinario:2020uvt} to the study of $p_T$-differential energy loss of jets in PbPb collisions. To that end, we focus our analysis on Z+jet events.

In general, the production of jets tagged by photons and electroweak bosons has been regarded as a `golden channel' for jet quenching studies~\cite{Brewer:2021hmh}, as these particles do not interact strongly with the QGP. This allows to approximate the initial $p_{T}$ of the jets (or rather, of the partons that originate the jets) with that of the measured trigger bosons. Such a strategy conveniently avoids the biases that dominate $p_T$-differential jet observables in other event topologies (such as dijet events~\cite{CMS:2013lhm,ATLAS:2018bvp,ATLAS:2010isq,CMS:2011iwn}) and provides a straightforward measurement of energy loss through the momentum imbalance between jet and Z-boson~\cite{CMS:2021otx,ATLAS:2019dsv,CMS:2017ehl,CMS:2017eqd,CMS:2018jco,ATLAS:2018dgb}. Another benefit of Z-tagging is that the samples obtained through this technique are dominated by quark-initiated jets~\cite{ATLAS:2019dsv}, a constraint that reduces the uncertainties introduced by potentially flavor-dependent effects in energy loss. 

While previous efforts have to a large extent focused on reducing the selection bias and obtaining jet samples with different quark and gluon fractions the prime goal of working with the $\tau$ algorithm (and $\tau_{\text{form}}$) is to systematically explore to what extent information about the temporal structure of jets and the QGP can be extracted from reconstructed jets. For instance, selecting on formation time allows us to construct jet populations that are more or less susceptible to energy loss while keeping the selection bias minimal. We here demonstrate how this procedure works in $Z$+jet events where the selection bias can be quantified by looking at the Z-boson $p_T$-spectrum. We would, however, like to point out that the same method can be applied to single-inclusive jets or dijets.

This paper is structured as follows. In \secref{sec:tau_alg} we briefly explain how to extract physical information from the substructure of jets by unclustering with the $\tau$ algorithm. We illustrate this novel technique by extracting \tauf~from Z-tagged jets, thereby classifying them according to their sensitivity to medium-induced energy loss. Then, in \secref{sec:tau_vs_deltaR} and \secref{sec:tau_vs_mg} respectively we compare \tauf- to \dR- and mass-based sampling as tools for jet quenching. Finally, we conclude and discuss future applications in \secref{sec:fin}.

\section{$\tau$ re-clustering algorithm for jet quenching studies}
\label{sec:tau_alg}

In this manuscript we focus our study on Z+jet $\sqrt{s_{NN}} = 5.02~\rm{TeV}$ events simulated with JEWEL~\cite{Zapp:2013vla,KunnawalkamElayavalli:2016ttl} v2.2.0. For the medium-induced effects, we restricted ourselves to the medium toy model provided within this event generator (Bjorken model~\cite{PhysRevD.27.140} for a boost-invariant longitudinal expansion of an ideal quark-gluon gas), with an initialisation time set to $\tau_i = 0.4 ~\rm{fm/c}$ and average initial temperature $T_i = 0.44~\rm{GeV}$. These values are known to provide an $R_{AA} = 0.4$, in agreement with current experimental observations at the same centre-of-mass energy~\cite{ALICE:2018vuu,ATLAS:2018gwx,CMS:2021vui}. The medium response is typically accounted for by including the recoiling scattering centers in the final state. However, in the present study we discard these particles in order to first study the effect of selecting jets based on formation time in a simpler and cleaner set-up. The extension including medium response effects will be discussed in an upcoming publication.

Following existing experimental analysis on boson-jet momentum asymmetry (e.g, CMS analysis~\cite{CMS:2017eqd}), the leading anti-k$_T$ jet is required to be within the pseudo-rapidity $|\eta_{j}| < 1.6$ and recoiling against a Z-boson with azimuthal distance $\Delta \phi = |\phi_{j} - \phi_Z| > 7\pi/8$. The Z-boson is reconstructed from the highest transverse momentum $\mu^+$ and $\mu^-$ pair with a minimum $p_{T} > 10~\rm{GeV/c}$ and pseudo-rapidity $|\eta| < 2.4$, as employed in~\cite{CMS:2017eqd}. The Z-boson candidate is then accepted if the combined mass satisfies $m_{Z} \in [70, 110]~\rm{GeV/c^2}$. For the jet radius, we chose $R = 0.5$ to increase the probability of catching a larger angle (earlier) emission within the jet area, as previously demonstrated in~\cite{Apolinario:2020uvt}\footnote{Increasing the jet radius also results in an enhanced sensitivity to soft contamination; however, the value $R=0.5$ was found to achieve a balance between this undesired effect and the improvement of the initial jet energy reconstruction.}. Moreover, we also require a minimum transverse momentum of $p_{T,Z}^{min} = 60~\rm{GeV/c}$ for the boson and $p_{T,j}^{min} = 30~\rm{GeV/c}$ for the reconstructed away-side jet. Once the pair of objects is identified, we follow the same jet substructure analysis as proposed in~\cite{Apolinario:2020uvt}. The jet constituents are re-clustered with the $\tau$ algorithm (generalized $k_T$ algorithm with $p = 0.5$) to obtain a sequence of clustering steps ordered in inverse formation time. We then select the first unclustering step to survive the Soft-Drop (SD) condition\footnote{Less than 1\% of the events within the chosen kinematic region is discarded when requiring jets to have at least one emission that satisfies \eqref{eq:sd}.}.:
\begin{equation}
    z_g = \frac{\text{min}(p_{T,1},p_{T,2})}{p_{T,1}+p_{T,2}}>z_\text{cut} =0.1 \, ,
    \label{eq:sd}
\end{equation}
where $p_{T,1(2)}$ is the transverse momentum of the (sub-)leading subjet obtained from the unclustering. These analysis steps are performed within FastJet v3.3.0~\cite{Cacciari:2011ma}.

Once identified the first emission from a $\tau$-declustered tree that satisfies \eqref{eq:sd}, we estimate the \textit{formation time}~\cite{Dokshitzer:1991wu} as:
\begin{equation}
    \tauff \simeq \frac{1}{2 E z_1 z_2 (1-\cos\theta_{12})} \, ,
    \label{eq:tauf}
\end{equation}
where $z_{1(2)}$ is the energy fraction carried by the (sub-)leading subjet, $E$ the total jet energy and $\theta_{12}$ the angle obtained using the three-dimensional momentum vectors of the two subjets. In fact, in~\cite{Apolinario:2020uvt}, it was shown that the unclustering with the $\tau$ algorithm provides a good correlation between the \tauf~obtained via unclustering and the one within the parton shower for each splitting along the primary branch. In this work we focus on the first emission only, leaving possible uses of subsequent unclustering steps for future work.

In addition, we compute the boson-jet momentum asymmetry:
\begin{equation}
    x_{j,Z}=\frac{p_{T,j}}{p_{T,Z}} \, ,
    \label{eq:xjz}
\end{equation}
as an estimate of energy loss effects induced by the presence of a QGP since the transverse momentum of the Z-boson can be used, to first order, as a good proxy for the jet-initiating parton energy. This correspondence, although deteriorated by initial state radiation, is appropriate insofar as we limit ourselves to using the overall distribution. In this case, it is enough for the correlation between $p_{T,\text{parton}}$ and $p_{T,Z}$ to be approximately the same in pp and AA collisions, which we expect to be a fair assumption within this level of precision. The correlation of the transverse momenta of the Z-boson and the jet-initiating parton depends only on the hard scattering and processes happening in the initial state. Therefore, only nuclear effects from parton distribution functions can additionally contribute to a smearing in AA collisions, but these are expected to be of moderate size and not significantly distort the correlation in these systems.

\begin{figure}
\centering
  \includegraphics[width=.9\linewidth]{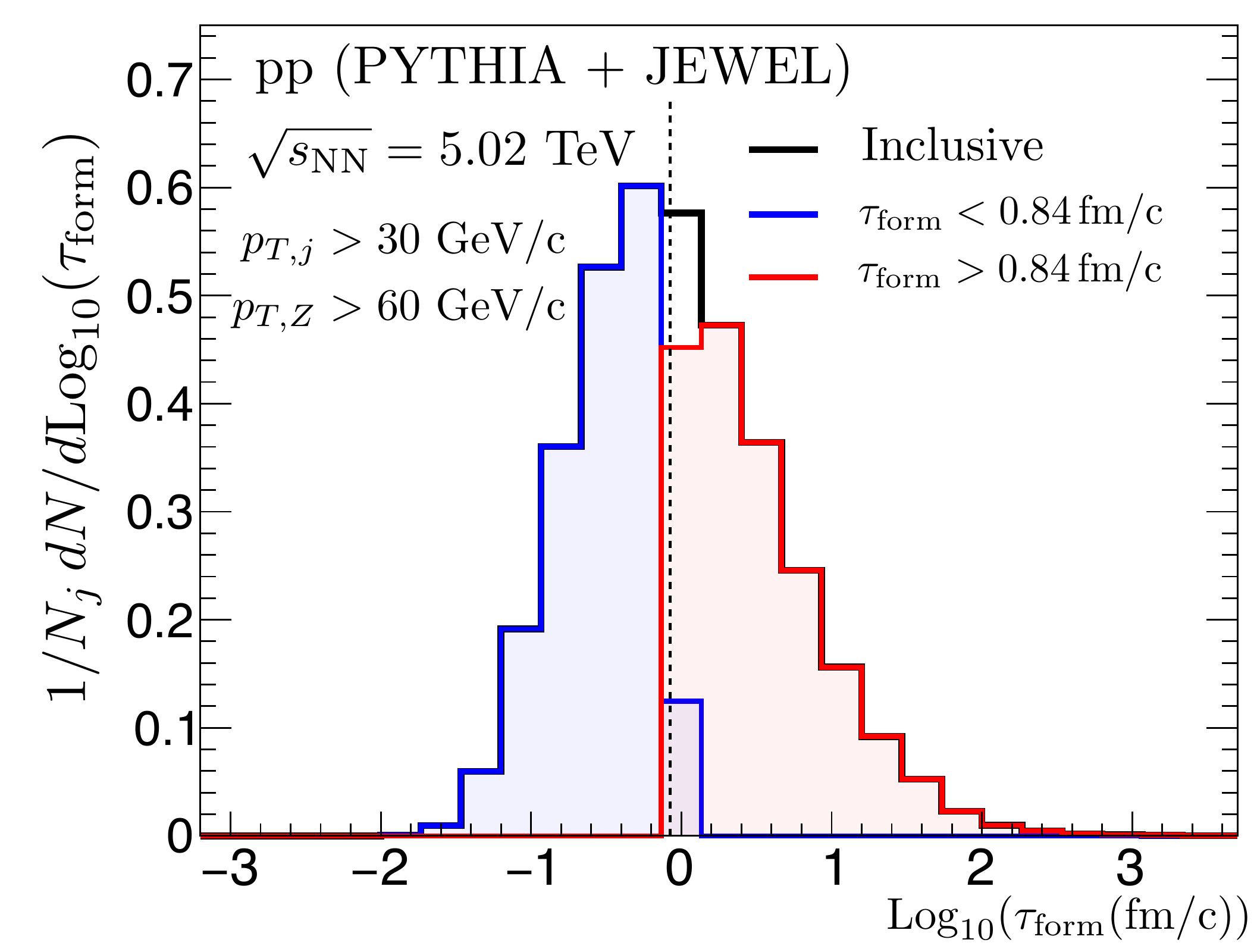}
  \includegraphics[width=.9\linewidth]{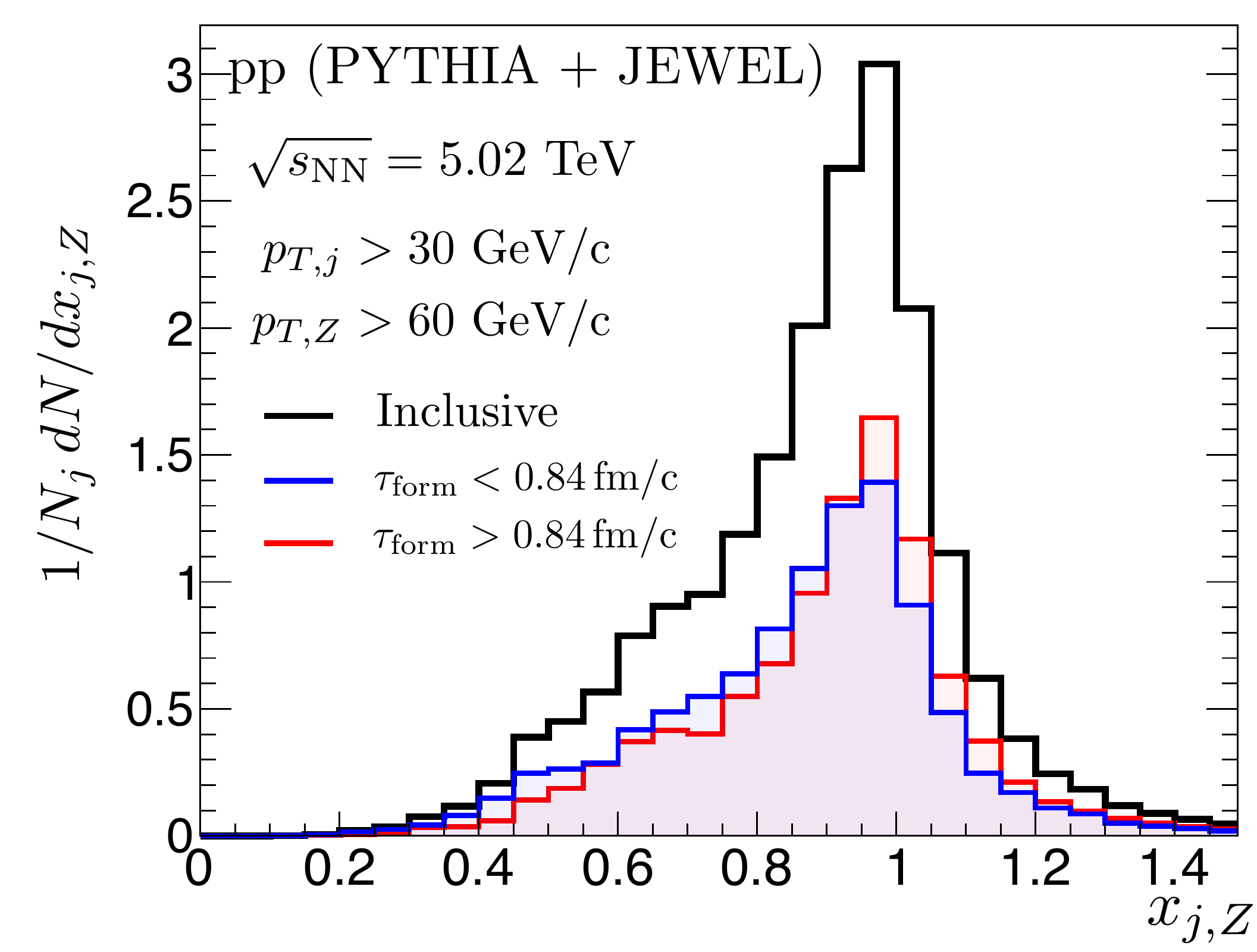}
\caption{Formation time (\tauf) of the first $\tau$-declustering step that satisfies $z_g>z_\text{cut}$ (top panel) and boson-jet momentum asymmetry (\xjz, bottom panel) distributions obtained from $\sqrt{s_{NN}} = 5.02~\rm{TeV}$ Z+jet JEWEL+PYTHIA pp events, normalized to the total number of jets in the sample. The blue distribution refers to the $50\%$ of jets that have a smaller formation time, while the red distribution characterizes the $50\%$ of jets that have a larger formation time. The median of \tauf~in this sample yields $0.84~\rm{fm/c}$.}
\label{fig:tau_bins_pp}
\end{figure}

\begin{figure}
\centering
  \includegraphics[width=.9\linewidth]{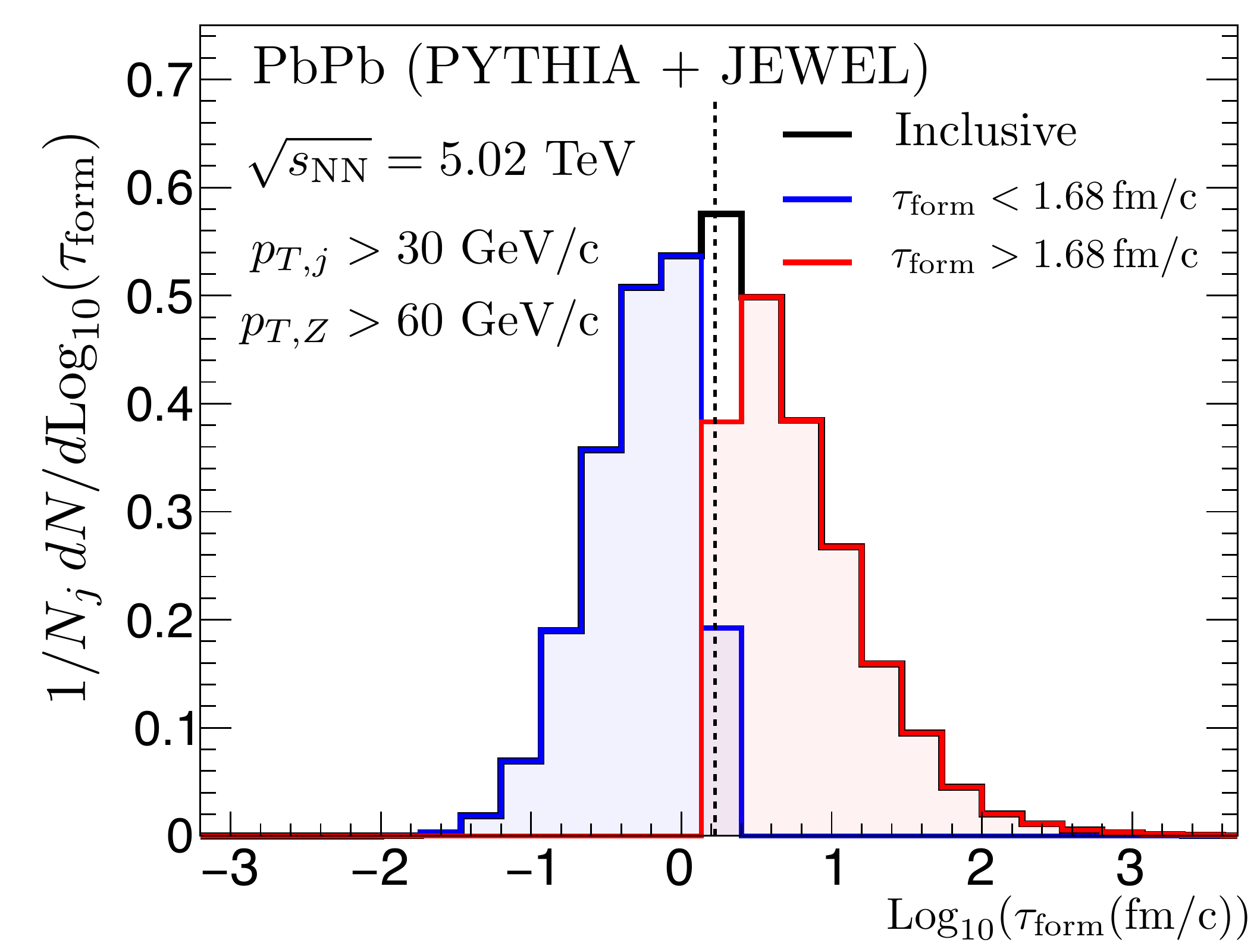}
  \includegraphics[width=.9\linewidth]{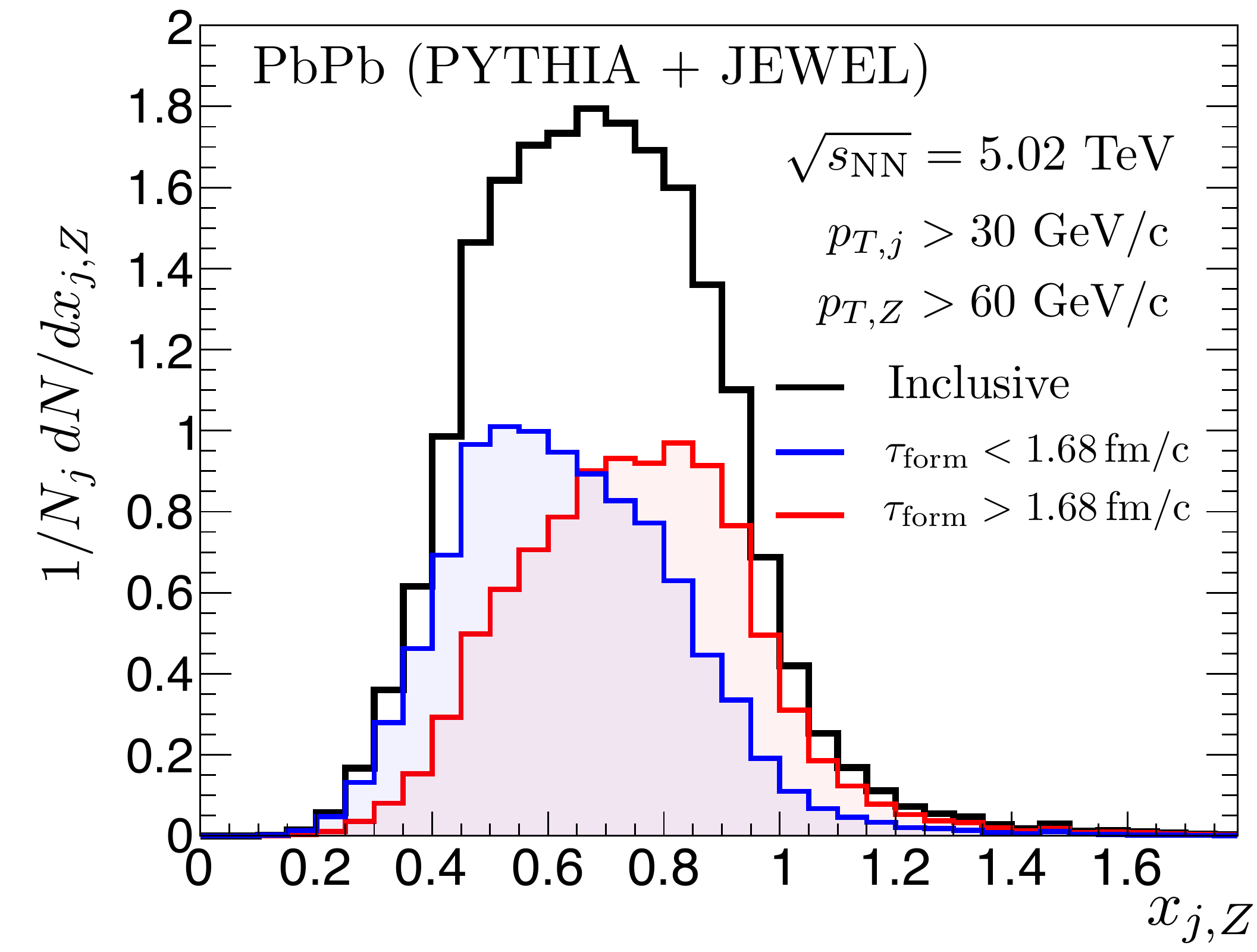}
\caption{Formation time (\tauf) of the first $\tau$-declustering step that satisfies $z_g>z_\text{cut}$ (top panel) and boson-jet momentum asymmetry (\xjz, bottom panel) distributions obtained from $\sqrt{s_{NN}} = 5.02~\rm{TeV}$ Z+jet JEWEL+PYTHIA PbPb events, normalized to the total number of jets in the sample. The blue distribution refers to the $50\%$ of jets that have a smaller formation time, while the red distribution characterizes the $50\%$ of jets that have a larger formation time. The median of \tauf~in this sample yields $1.68~\rm{fm/c}$.}
\label{fig:tau_bins_PbPb}
\end{figure}

The distributions of the formation time (\tauf) and boson-jet momentum asymmetry (\xjz) of the events without medium-induced effects (JEWEL + PYTHIA pp) are shown in the top and bottom panel of \figref{fig:tau_bins_pp} respectively, in black solid lines. The corresponding distributions when medium-induced effects are included (JEWEL + PYTHIA PbPb) are shown in \figref{fig:tau_bins_PbPb}. We also display the populations of jets below and above the median value of each of the \tauf~distributions in blue and red respectively. Such classes will be designated as \textit{early-} or \textit{late-}initiated jets. This classification was previously found to strongly distinguish between different degrees of quenching, with the former having a nuclear modification factor, $R_{AA}$, smaller than unity, while the latter yields an $R_{AA} \simeq 1$. As such, it is no surprise to see a correlation between these populations of jets and the \xjz~distribution for PbPb collisions. Also here, early-initiated jets experience stronger medium modifications and thus have lower $x_{j,Z}$ than late-initiated jets. Importantly, there is only a weak correlation between the formation time of the first $\tau$-declustering step that satisfies $z_g>z_\text{cut}$ and the boson-jet momentum asymmetry in pp collisions. This opens the possibility to have the notion of a \textit{formation time} within the medium acting as a proxy for medium-induced energy loss. In the following, we will explore this further, while also presenting a comparison with a selection based on the radial distance between the two subjets and the groomed jet mass. 


\begin{figure}
\centering
  \includegraphics[width=.9\linewidth]{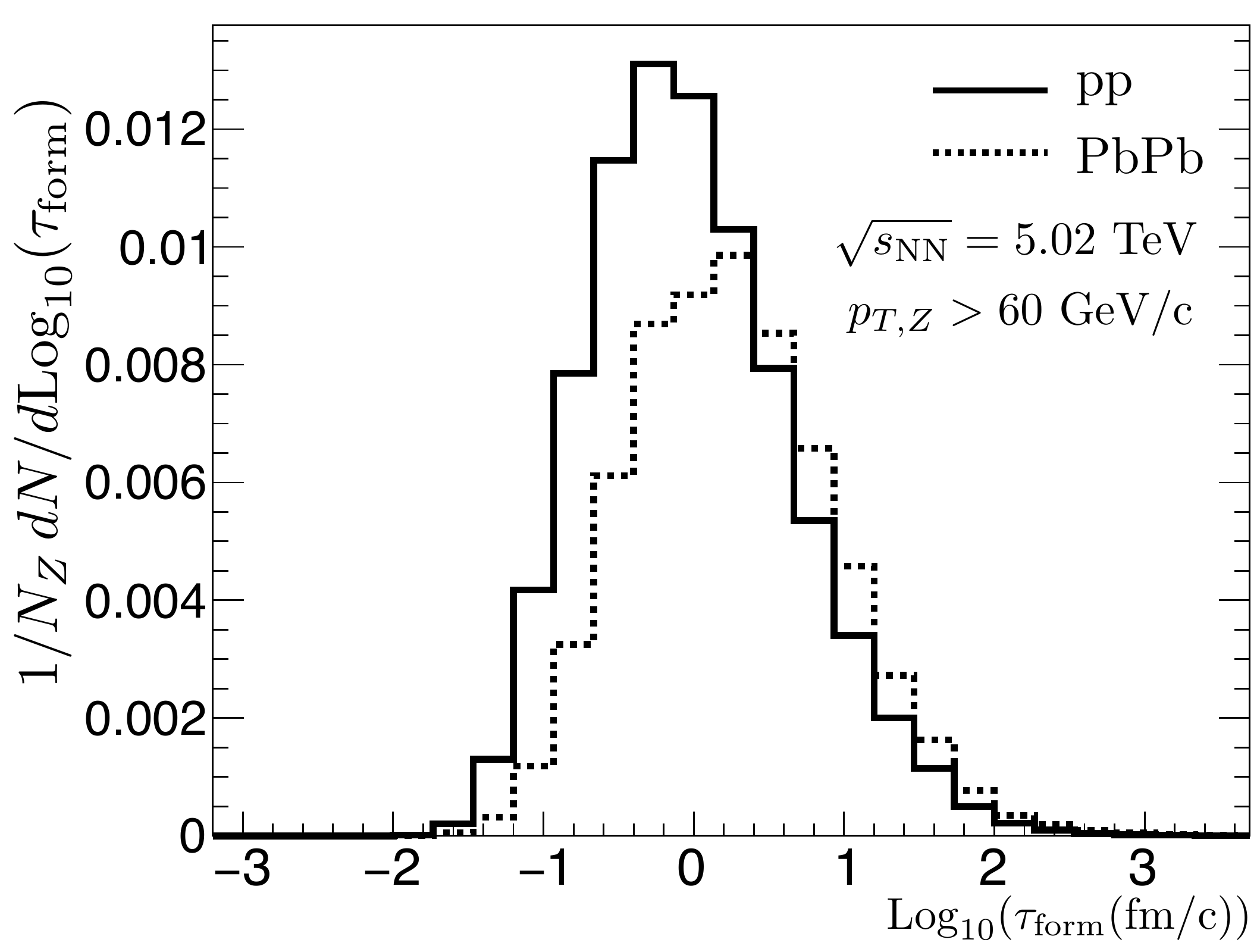}
\caption{\tauf~distributions for Z-tagged jets in pp and PbPb events, normalized to the total number of reconstructed Z-bosons (with or without accompanying jets).}
\label{fig:tau_dists}
\end{figure}

As a final note, when considering in-medium interactions, the reconstructed jet will have a smaller transverse momentum when compared to pp collisions, resulting in an overall shift of \xjz~towards smaller values. Simultaneously, the \tauf~distribution migrates towards larger values as clearly shown by the median of the distributions ($\tauff^{median} = 0.84~\rm{fm/c}$ for pp and $\tauff^{median} = 1.68~\rm{fm/c}$ for PbPb). This survival bias effect was previously observed in~\cite{Rajagopal:2016uip,Apolinario:2020uvt}. Jets with a smaller angle between the two subjets obtained via $\tau$-declustering with a $z_\text{cut}$, which typically have a larger \tauf~(see section~\ref{sec:tau_vs_deltaR}), have an increased probability of surviving the transverse momentum threshold in heavy-ion collisions. Conversely, wider jets, with typically shorter \tauf, tend to have a softer fragmentation pattern with more constituents and therefore lose more energy. They are thus reconstructed with a smaller $p_T$ with respect to pp collisions \footnote{The same effect is expected from colour coherence arguments where a small angle splitting remains unresolved by the medium while the daughters of a large angle splitting are resolved individually and lose energy incoherently. This effect is, however, not implemented in JEWEL.}. This effect is clearly shown in \figref{fig:tau_dists}, where, in solid, it is represented the \tauf~distribution of jets in Z+jet pp events, while in dashed the corresponding to PbPb events, both of them normalized to the total number of reconstructed Z-bosons.

\section{Comparison of \tauf~and \dR-based jet selection}\label{sec:tau_vs_deltaR}

In the previous section, we showed how the value of \tauf~corresponding to the first splitting that satisfies $z_g>z_\text{cut}$ can be used to define jet populations with enhanced sensitivity to medium-induced energy loss. This feature alone, however, is not unique to \tauf~and can be reproduced with almost identical results by performing a selection of jets based on the radial distance between the two obtained subjets~\cite{ALICE:2021mqf,Mehtar-Tani:2016aco,Chien:2016led}:
\begin{equation} \label{eq:deltaR}
\dRR = \sqrt{ (y_1 - y_2)^2 + (\phi_1 - \phi_2)^2 } \, ,
\end{equation}
with $y_{1(2)}$ and $\phi_{1(2)}$ being the rapidity and azimuthal angle of the (sub-)leading subjets. This correlation emerges from an implicit dependence of \tauf~ on \dR, which becomes apparent when re-writting \eqref{eq:tauf} with the transverse kinematics of the resulting subjets. Namely,
\begin{equation}\label{eq:tauf2}
    \tauff \simeq \frac{p_{T,1}+p_{T,2}}{p_{T,1}p_{T,2} \dRR^2} \, .
\end{equation}
Given that Eqs.~(\ref{eq:tauf}) and~(\ref{eq:tauf2}) do not significantly differ in the chosen kinematic region (see appendix \ref{app:tau_comp}), we will continue to use the definition of \eqref{eq:tauf} in the following.

\begin{figure}
\centering  \includegraphics[width=.9\linewidth]{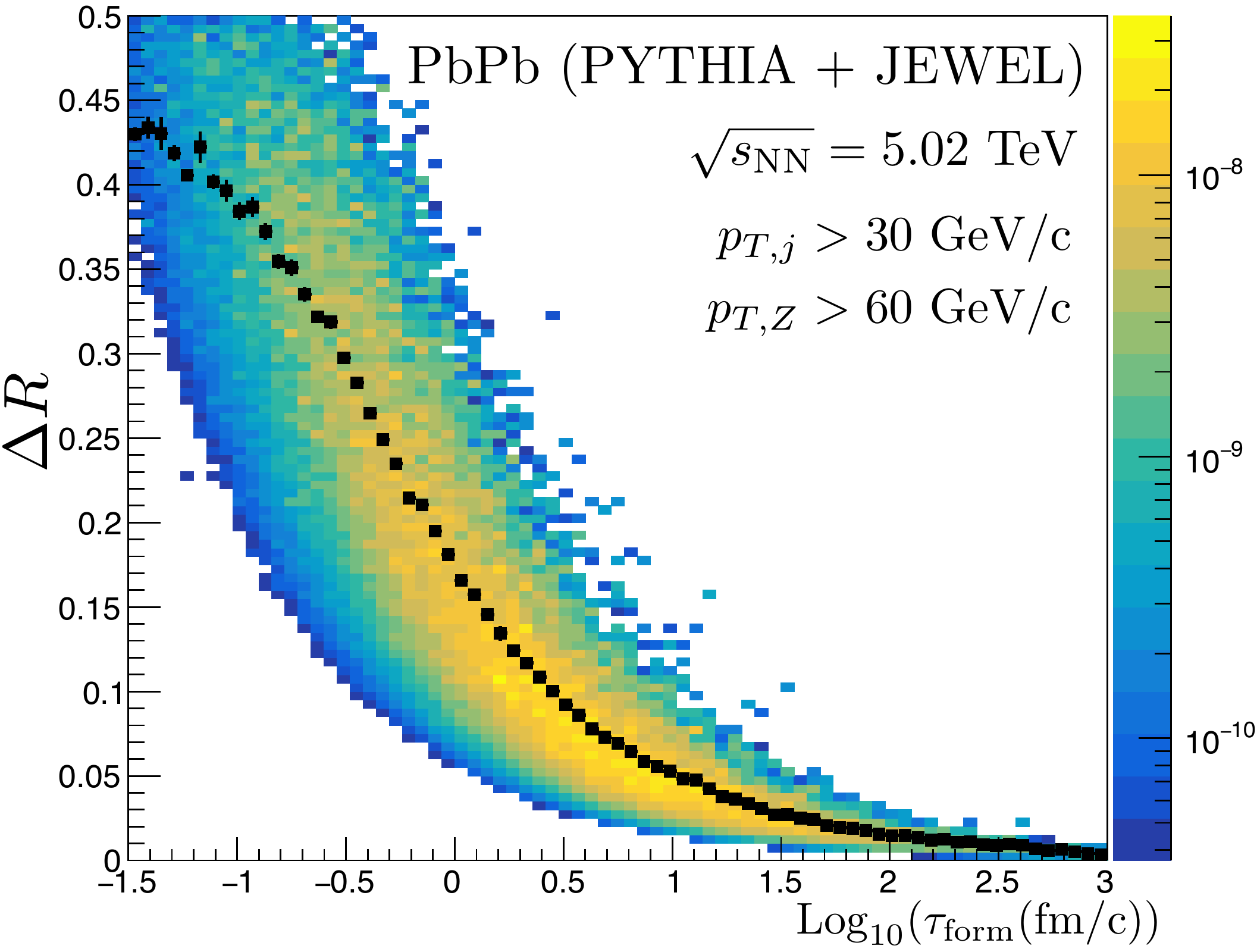}
\caption{Correlation between formation time (\tauf) and opening angle (\dR) of the first $\tau$-declustering step that satisfies $z_{g} >z_{cut}$ obtained from $\sqrt{s_{NN}} = 5.02~\rm{TeV}$ Z+jet JEWEL+PYTHIA PbPb events.}
\label{fig:correl_tau_deltaR}
\end{figure}

The correlation between \tauf~and \dR~for each jet is shown in \figref{fig:correl_tau_deltaR} for JEWEL+PYTHIA PbPb events (similar results are obtained for pp events). It is possible to observe that there is a strong correlation between the two variables, albeit with some degree of smearing. Small formation times typically correspond to large opening angles (\textit{first} emissions will have the largest opening angle) while small opening angles are strongly correlated with larger formation times. The question thus arises whether there is any benefit in selecting jets based on $\tauff$ instead of $\Delta R$. To answer this question, we established two different methodologies to inspect the jet substructure in our boson+jet simulated events. In one of them, we employ the $\tau$ algorithm to reconstruct the jet constituents and compute the \dR~from the first SD-approved (\eqref{eq:sd}) unclustering step. In the other, we instead adopt the more conventional approach of re-clustering with C/A~\cite{Dokshitzer:1997in,Wobisch:1998wt}, and apply the SD criteria to finally obtain \dR~of the two subjets. In the following, we will showcase only the results corresponding to $\tau$-re-clustered jets (first methodology) and refer the author to appendix \ref{app:ca_vs_tau} for a comparison with the C/A re-clustering.

\begin{figure}
\centering
  \includegraphics[width=.9\linewidth]{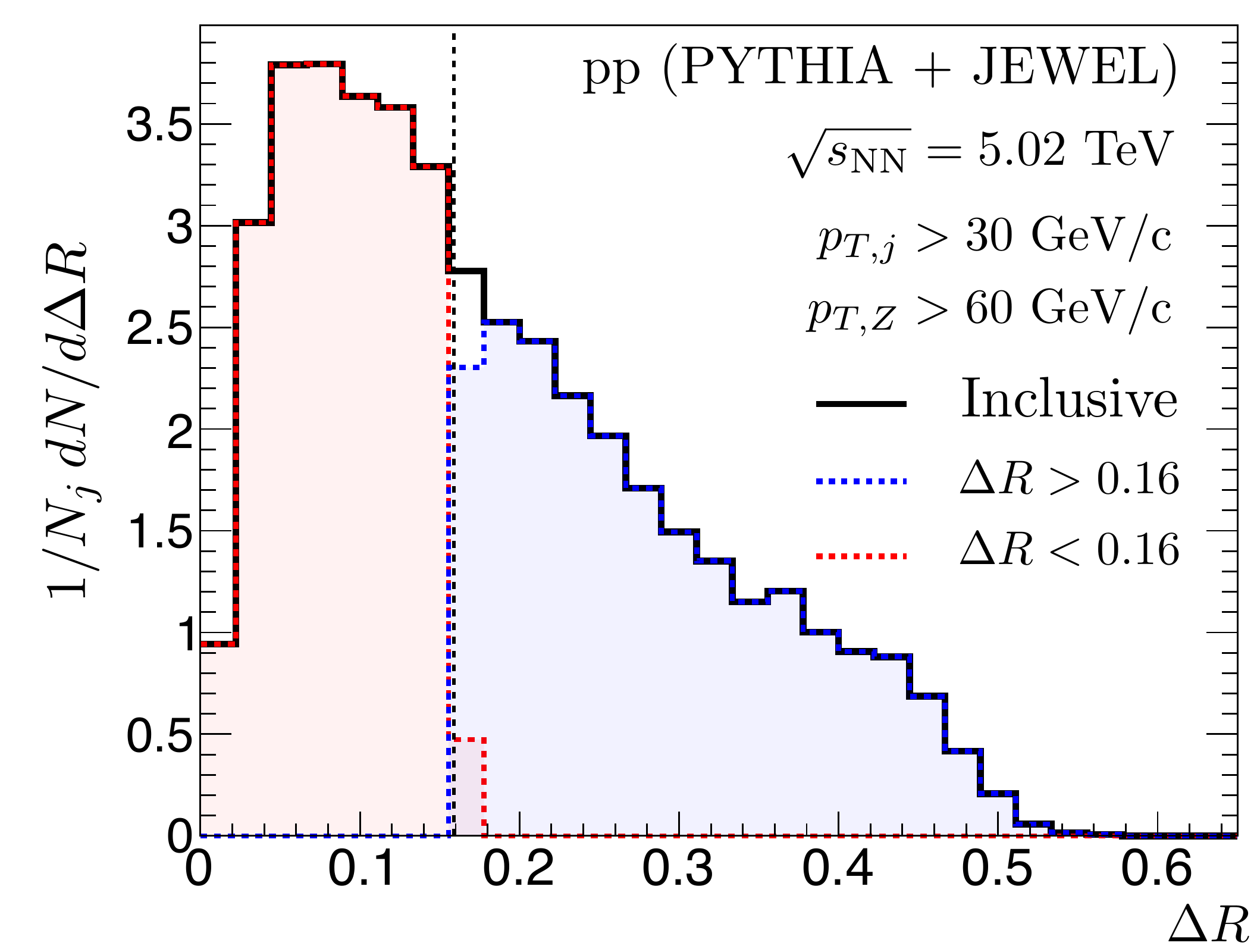}
  \includegraphics[width=.9\linewidth]{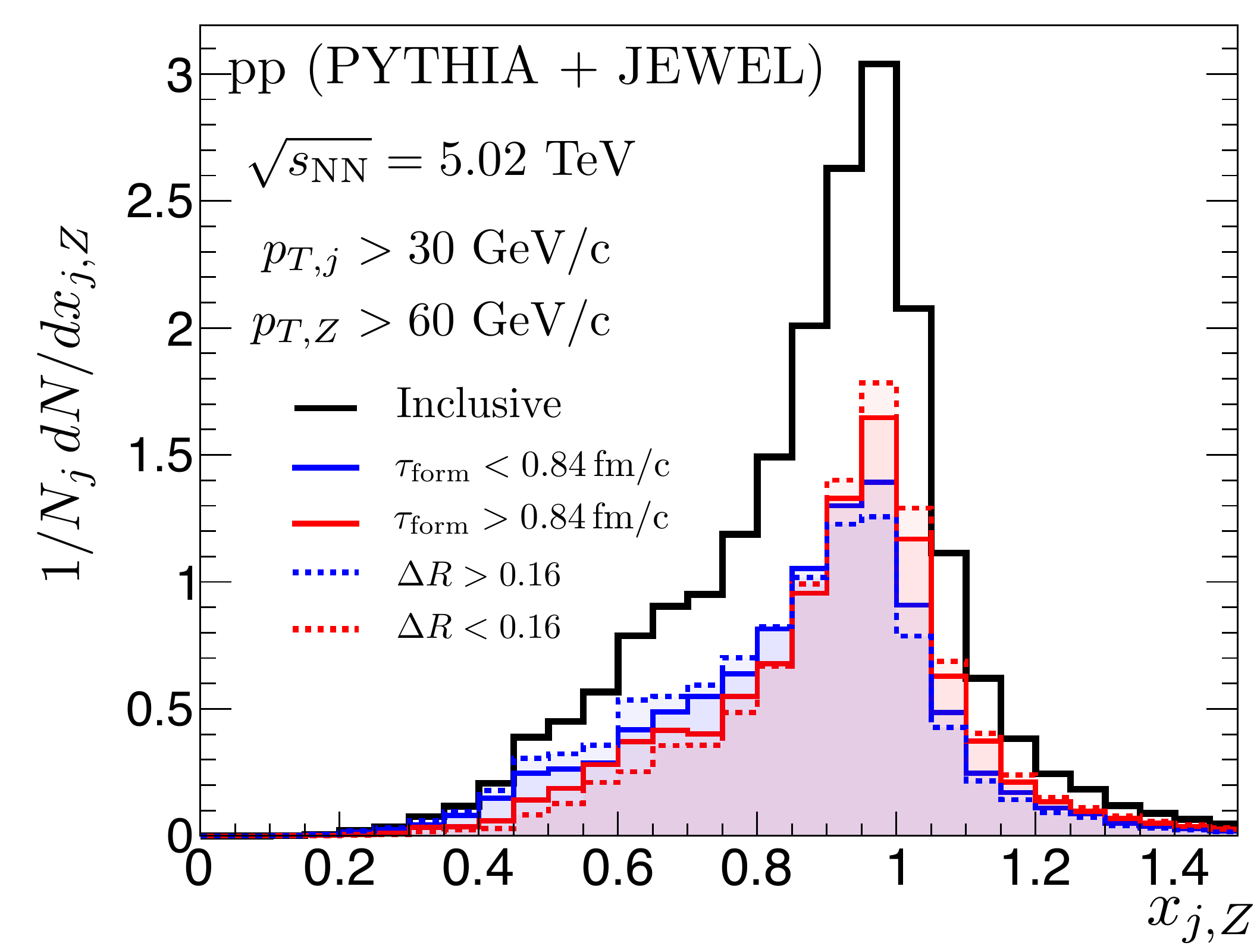}
\caption{Opening angle (\dR) of the first $\tau$-declustering step that satisfies $z_g>z_\text{cut}$ (top panel) and boson-jet momentum asymmetry (\xjz, bottom panel) distributions obtained from $\sqrt{s_{NN}} = 5.02~\rm{TeV}$ Z+jet JEWEL+PYTHIA pp events, normalized to the total number of jets in the sample. The blue distribution refers to the $50\%$ of jets that have a wider opening angle, while the red distribution characterizes the $50\%$ of jets that a smaller opening angle. The median of \dR~in this sample yields $0.16$. In the bottom panel we also show the distributions corresponding to selecting jets based on \tauf.}
\label{fig:dR_bins_pp}
\end{figure}

\begin{figure}
\centering
  \includegraphics[width=.9\linewidth]{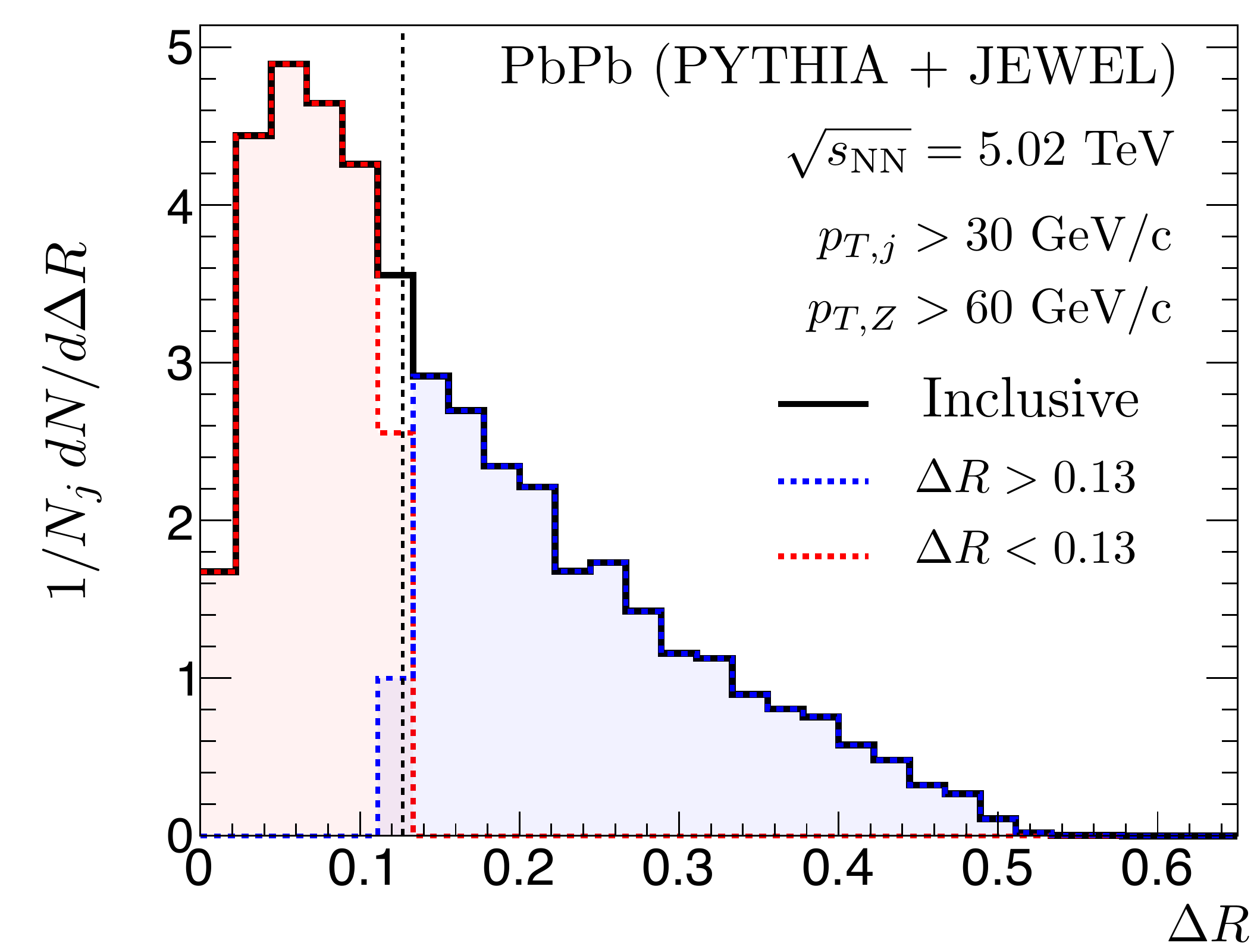}
  \includegraphics[width=.9\linewidth]{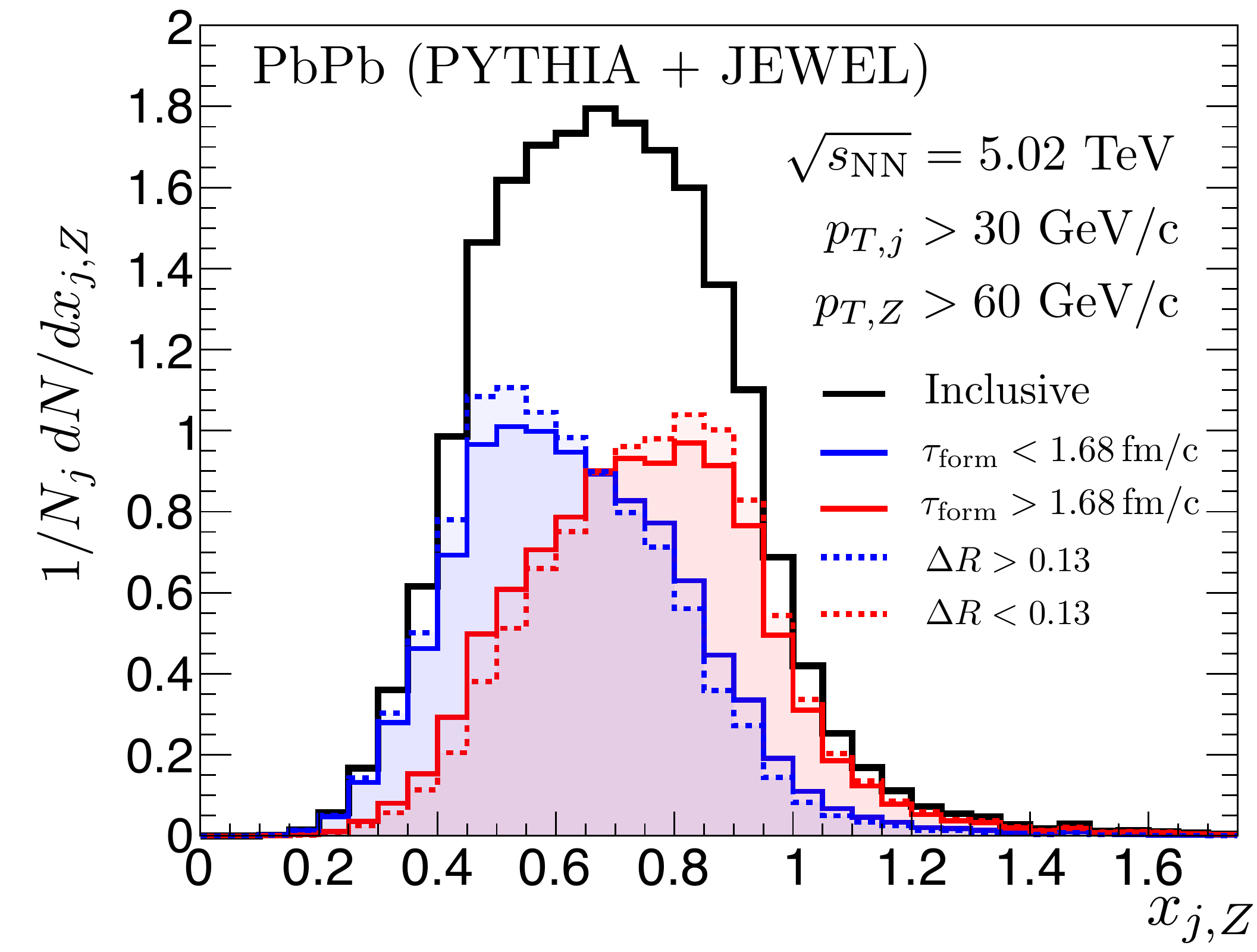}
\caption{Opening angle (\dR) of the first $\tau$-declustering step that satisfies $z_g>z_\text{cut}$  (top panel) and boson-jet momentum asymmetry (\xjz, bottom panel) distributions obtained from $\sqrt{s_{NN}} = 5.02~\rm{TeV}$ Z+jet JEWEL+PYTHIA PbPb events, normalized to the total number of jets in the sample. The blue distribution refers to the $50\%$ of jets that have a wider opening angle, while the red distribution characterizes the $50\%$ of jets that a smaller opening angle. The median of \dR~in this sample yields $0.13$. In the bottom panel we also show the distributions corresponding to selecting jets based on \tauf.}
\label{fig:dR_bins_PbPb}
\end{figure}

In order to investigate the differences between using \dR~or \tauf~as a selector for quenching, we will adopt the same procedure of splitting the jet sample into half, but selecting, in this case, the $50\%$ narrower and $50\%$ wider samples of the total available jet populations according to their median (0.16 for pp and 0.13 for PbPb). Thus, the equivalent of Figs.\ \ref{fig:tau_bins_pp} and \ref{fig:tau_bins_PbPb} but now using a selection based on \dR~is shown in Figs.\ \ref{fig:dR_bins_pp} and \ref{fig:dR_bins_PbPb}. 

The top panels illustrate the overall \dR~distributions obtained for pp and PbPb events, with the dashed red and blue selections representing the $50\%$ narrower and $50\%$ wider samples of the total available jet populations, respectively. In the bottom panels, we show the contributions of each sub-sample to the overall \xjz~distributions in both pp and PbPb, while simultaneously showing the \textit{early} (blue solid line) and \textit{late} (red solid line) jet selections defined in section~\ref{sec:tau_alg}. Selecting jets based on \dR~yields qualitatively the same populations as selecting on \tauf. PbPb jets with a collinear pattern (dashed red line) show a smaller \xjz~shift with respect to the pp reference than those with a wide opening angle (dashed blue line) and these fairly correspond to the same \textit{late} and \textit{early} jet populations as defined in section~\ref{sec:tau_alg}. 

\begin{figure}
\centering
  \includegraphics[width=.9\linewidth]{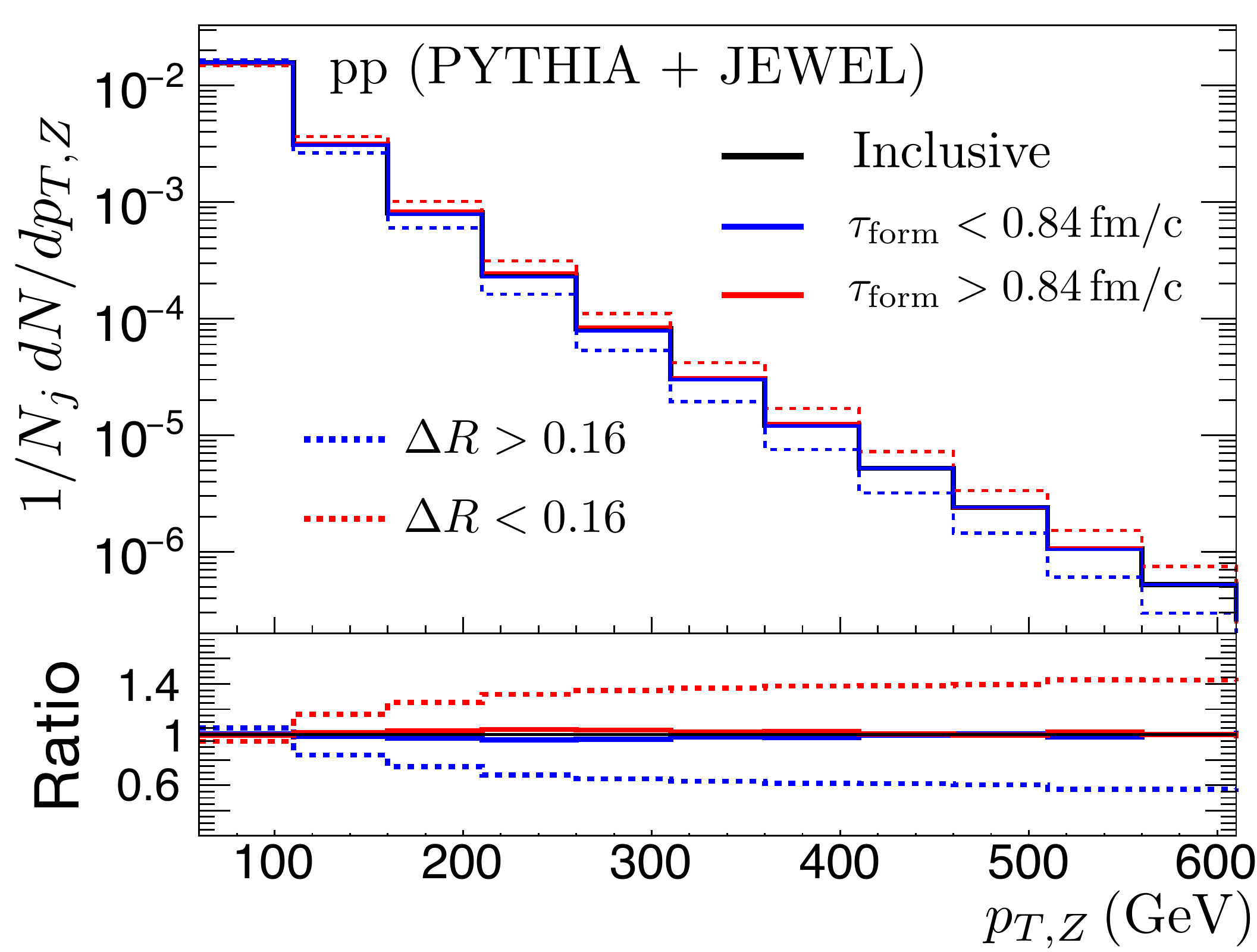}
  \includegraphics[width=.9\linewidth]{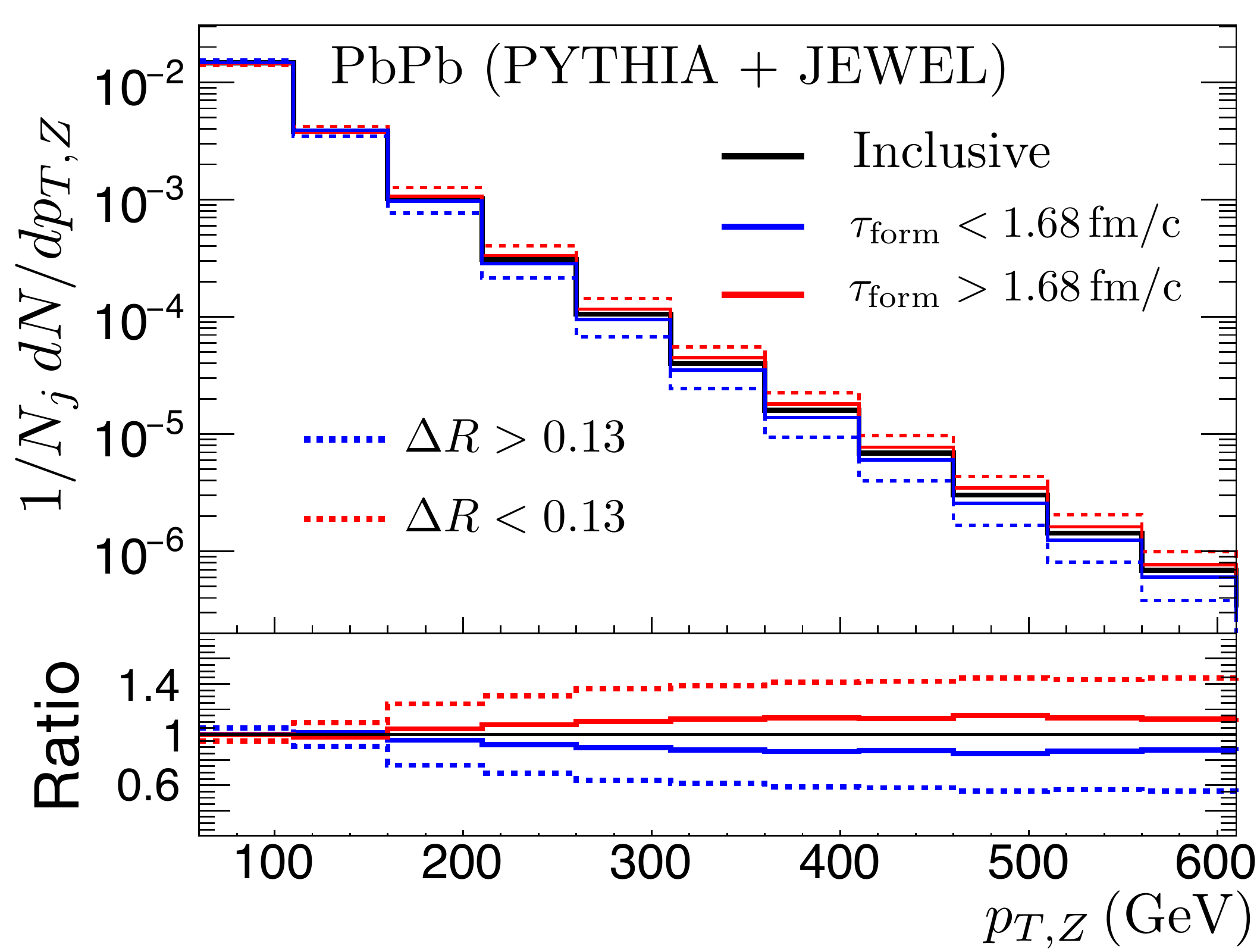}
\caption{Comparison of $p_{T,Z}$ spectra for jet samples defined for $\sqrt{s_{NN}} = 5.02~\rm{TeV}$ Z+jet JEWEL+PYTHIA pp (top panel) and PbPb (bottom panel) events, normalized to the size of each sample. The solid curves represent the early (blue) and late (red) jet spectra, defined according to their values of \tauf, whereas the dashed curves represent the wide (blue) and narrow (red) jets, defined according to their values of \dR. Each of these values extracted from the first $\tau$-declustering step that satisfies the SD condition. The inclusive spectrum is shown in solid black. The bottom part of each panel displays the ratio between each selection and the inclusive distribution.}
\label{fig:bias_plots_50}
\end{figure}

This suggests that both \tauf- and \dR-based selections have a similar sensitivity to energy loss. In addition, taking into consideration the results from pp collisions, both seem to introduce the same kinematic bias as the resulting selection on the $x_{jZ}$ distribution does not align with the inclusive result. However, $x_{jZ}$ depends both on the (reconstructed) jet and boson momentum distribution. It is thus natural that the distribution changes when different jet populations are selected. The more relevant question is to what extent the distribution of jet-initiating partons is affected by the selection. To quantify this and to finally understand the differences between the two selection methods, we looked at the Z-boson transverse momentum spectra ($p_{T,Z}$) that correspond to each of these populations. The results are shown in \figref{fig:bias_plots_50} for pp (top panel) and PbPb events (bottom panel). The top parts of these plots show the inclusive $p_{T,Z}$ spectrum in black, the narrow (wide) jets as defined above in dashed red (blue) and the \textit{late} (\textit{early}) jets as defined in section~\ref{sec:tau_alg} in solid red (blue). The bottom part shows the ratio with respect to the inclusive result.

As previously mentioned, the transverse momentum of the Z-boson, whose distribution we show in \figref{fig:bias_plots_50}, is a proxy for the transverse momentum of the jet-initiating parton. The top panel illustrates pp collisions, where energy loss effects are absent. Despite the similarities in $x_{j,Z}$ distributions when selecting based on $\Delta R$ or $\tau_{\text{form}}$, the resulting event samples differ significantly. Jets characterized by a large $\Delta R$ (indicated by the blue dashed line) typically exhibit a correlation with softer initial $p_{T}$ values, leading to suppression at high $p_{T,Z}$. Conversely, jets with a smaller $\Delta R$ (shown by the red dashed line) tend to be associated with harder initial $p_{T}$ values, resulting in enhancement at high $p_{T,Z}$. However, such biases are markedly reduced when selecting based on $\tau_{\text{form}}$, as indicated by the solid lines. Consequently, the observed modifications in the $x_{j,Z}$ distribution are primarily driven by changes in the final reconstructed jet momenta. Nonetheless, for $\Delta R$ classes, the modifications represent a convolution between changes in the final reconstructed jet spectra and kinematic biases. Upon incorporating energy loss effects (as observed in PbPb collisions, top panel of \figref{fig:bias_plots_50}), both variables appear to introduce a bias in the survival sample of boson-jet pairs. However, $\tau_{\text{form}}$ minimizes this bias compared to $\Delta R$, thus providing a more transparent connection between the $x_{j,Z}$ distribution and QGP-induced energy loss. It is important to note that we anticipate an additional kinematic bias in PbPb collisions compared to pp: the surviving jets at the minimum transverse momentum threshold are typically those with harder fragmentation. This is inherent to the analysis and is present for both selectors. Since \tauf~goes with $\Delta R^{-2}$ and linear with $p_T$ (see equation \eqref{eq:tauf2}), some of the correlation between the transverse momentum of the initiating particle and the boosted distance between the two daughters is cancelled in \tauf. Naturally, this bias will increase with smaller \dR~and/or \tauf~jet selections. To illustrate the limitations of such selections, in appendix~\ref{app:bias_smaller_selections} we display samples containing $25\%$ of the total jet population. We would like to point out that this way of quantifying the selection bias relies solely on the observable final state particles and can therefore also be carried out on data. Therefore, we argue that a defining feature of the $\tau_{\text{form}}$, withdrawn with the $\tau$ algorithm, is that it allows to interpret the result unambiguously in terms of an internal scale of the parton shower (i.e.\ the time of the first SD-approved splitting within the medium), whereas the same cannot be said for the \dR-based classification.

Finally, it is worth reiterating that all \tauf~and \dR~values included in this work were extracted from the first unclustering step given by the $\tau$ algorithm (except where specifically indicated otherwise). Nevertheless the conclusions remain qualitatively the same, when C/A is used instead (see appendix~\ref{app:ca_vs_tau}) since the chosen procedure to split the jet population, and analyse possible bias, is based on the median of the obtained distributions. The absolute values of both \dR~and \tauf~change depending on the re-clustering algorithm, and, as noted before in ~\cite{Apolinario:2020uvt}, the $\tau$ algorithm provides the best methodology to estimate the \tauf~from the parton shower Monte Carlo-truth. As such, the values of the obtained medians will depend on the chosen re-clustering algorithm. However, the population of jets classified as early/late or narrow/wide are mostly the same, leading to marginally noticeable differences (see appendix~\ref{app:ca_vs_tau}). Thus, our choice of fixing the re-clustering algorithm to $\tau$ across this exercise.

\section{Comparison of \tauf~and $\rho_g$-based jet selection}\label{sec:tau_vs_mg}

As a final study, we investigate whether the (groomed) jet mass~\cite{CMS:2018fof} could be used instead of the formation time to make jet selections sensitive to the in-medium energy loss. Such a possibility is suggested by \eqref{eq:tauf}, which contains an explicit dependence on the total jet energy (and hence on the mass of the two subjets). The correlation becomes further evident when considering that our estimate of \tauf~consists simply in the lifetime of the parent quark on its rest frame (approximated as the inverse of its virtuality $Q$, i.e.\ its inverse invariant mass) boosted to the lab system by a factor $E/Q$.

\begin{figure}
\centering
  \includegraphics[width=.9\linewidth]{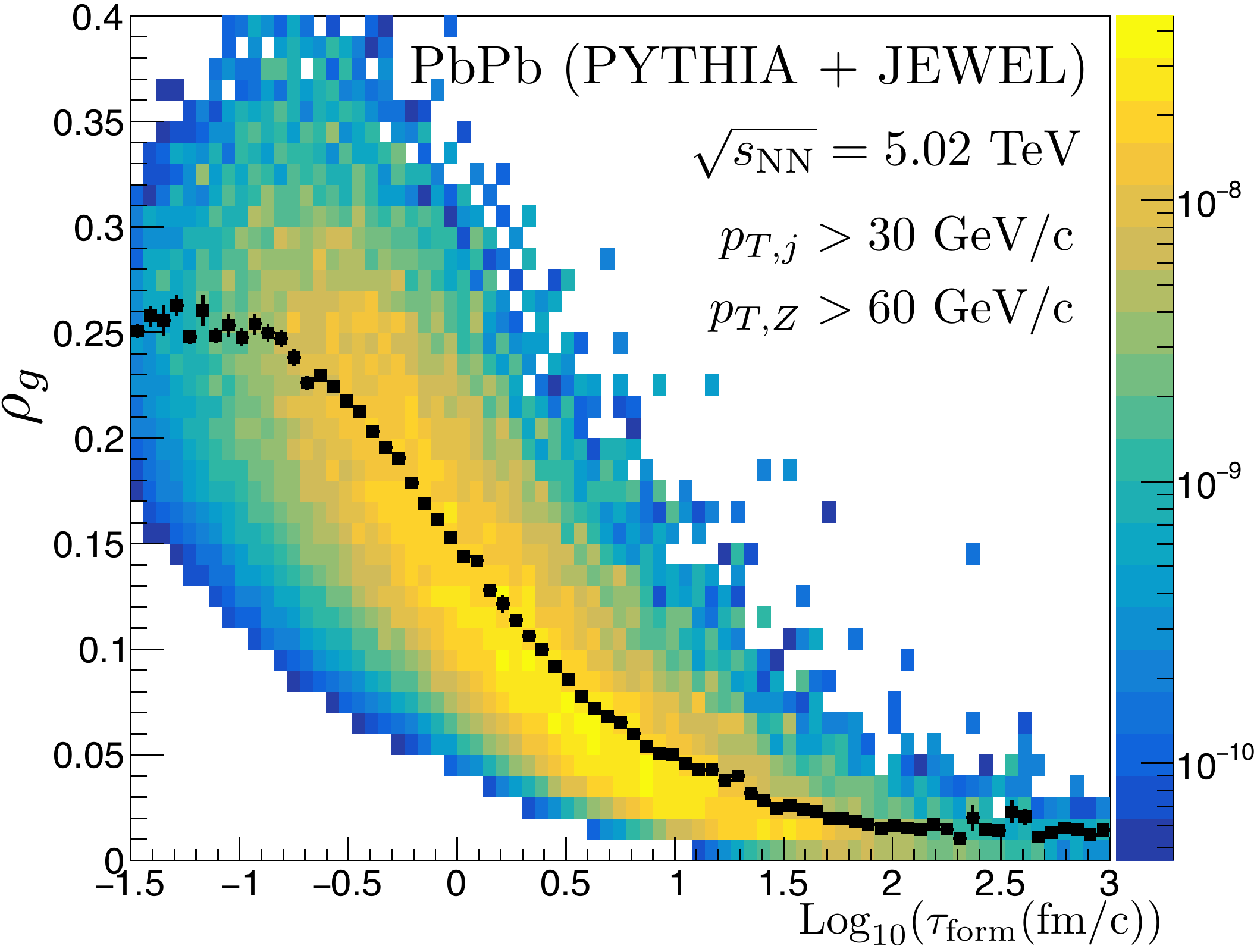}
\caption{Correlation between formation time (\tauf) and groomed jet mass ($\rho_g$) of the first $\tau$-declustering step that satisfies the SD condition, obtained from $\sqrt{s_{NN}} = 5.02~\rm{TeV}$ Z+jet JEWEL+PYTHIA PbPb events.}
\label{fig:correl_tau_mg}
\end{figure}

We focus our discussion on the groomed jet mass (computed as the sum of the masses of the subjets resulting from $\tau$-declustering with a $z_\text{cut}$) normalized by the ungroomed transverse momentum, $\rho_g\equiv m_g/p_{T,j}$. Both the mean and overall shape of the $\rho_g$ distribution are highly dependent on the clustering algorithm employed to obtain it (see appendix ~\ref{app:grooming}, \figref{fig:algorithm_comparison1_b}). This is in contrast with \tauf~and \dR, whose distributions show significantly less variation when re-clustering with different algorithms ($p\!=\!0$ and $p\!=\!0.5$) (see appendix~\ref{app:ca_vs_tau}). Nevertheless, we find that the apparent sensitivity of $\rho_g$ to the algorithm exponent $p$ does not propagate to the results discussed here. In fact, as with the \tauf~and \dR-based selections, we obtain very similar results for both algorithms considered given that we are focusing in the median of the obtained distributions rather than its absolute value. We will thus continue to show the results corresponding to the $\tau$ algorithm only, leaving a comparison with C/A for appendix~\ref{app:ca_vs_tau} (see also appendix~\ref{app:grooming} for a more detailed discussion of the grooming procedure).

\begin{figure}
\centering
  \includegraphics[width=.9\linewidth]{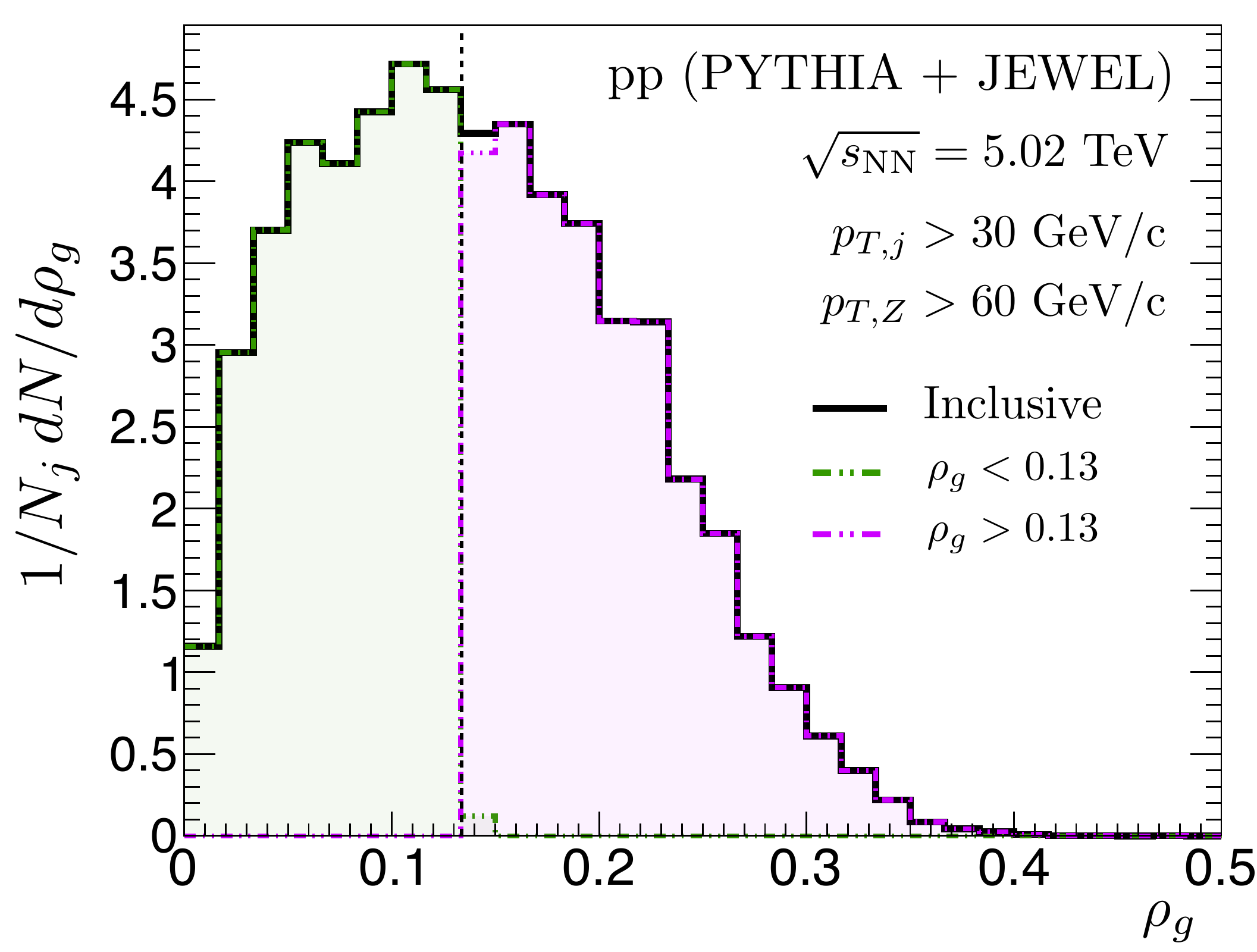}
  \includegraphics[width=.9\linewidth]{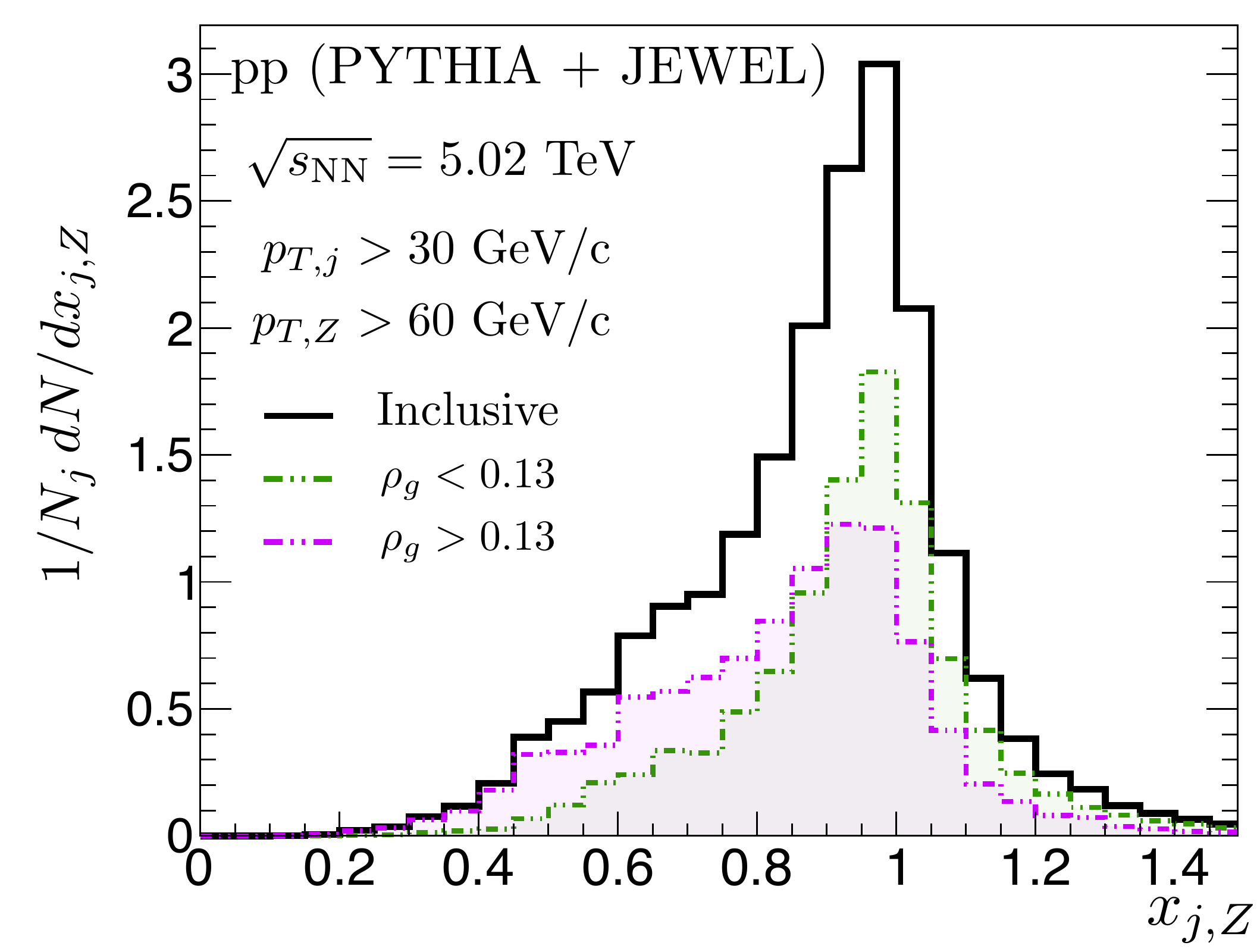}
\caption{Groomed jet mass ($\rho_g$) of the first $\tau$-declustering step that satisfies $z_g>z_\text{cut}$ (top panel) and boson-jet momentum asymmetry (\xjz, bottom panel) distributions obtained from $\sqrt{s_{NN}} = 5.02~\rm{TeV}$ Z+jet JEWEL+PYTHIA pp events, normalized to the total number of jets in the sample. The violet distribution refers to the $50\%$ heavier jets, while the green distribution characterizes the $50\%$ lighter jets. The median of $\rho_g$ in this sample yields 0.13.}
\label{fig:mg_bins_pp}
\end{figure}

\begin{figure}
\centering
  \includegraphics[width=.9\linewidth]{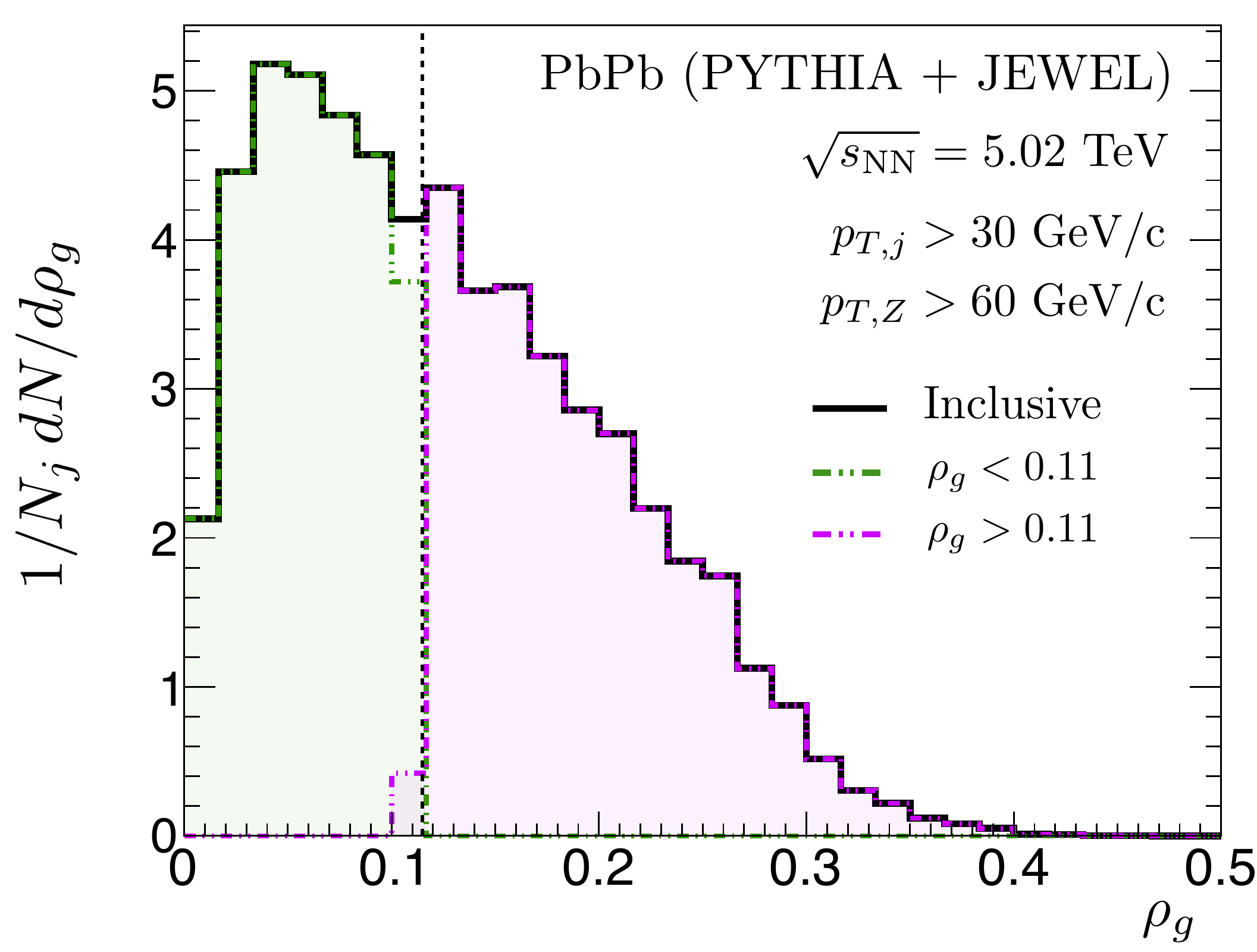}
  \includegraphics[width=.9\linewidth]{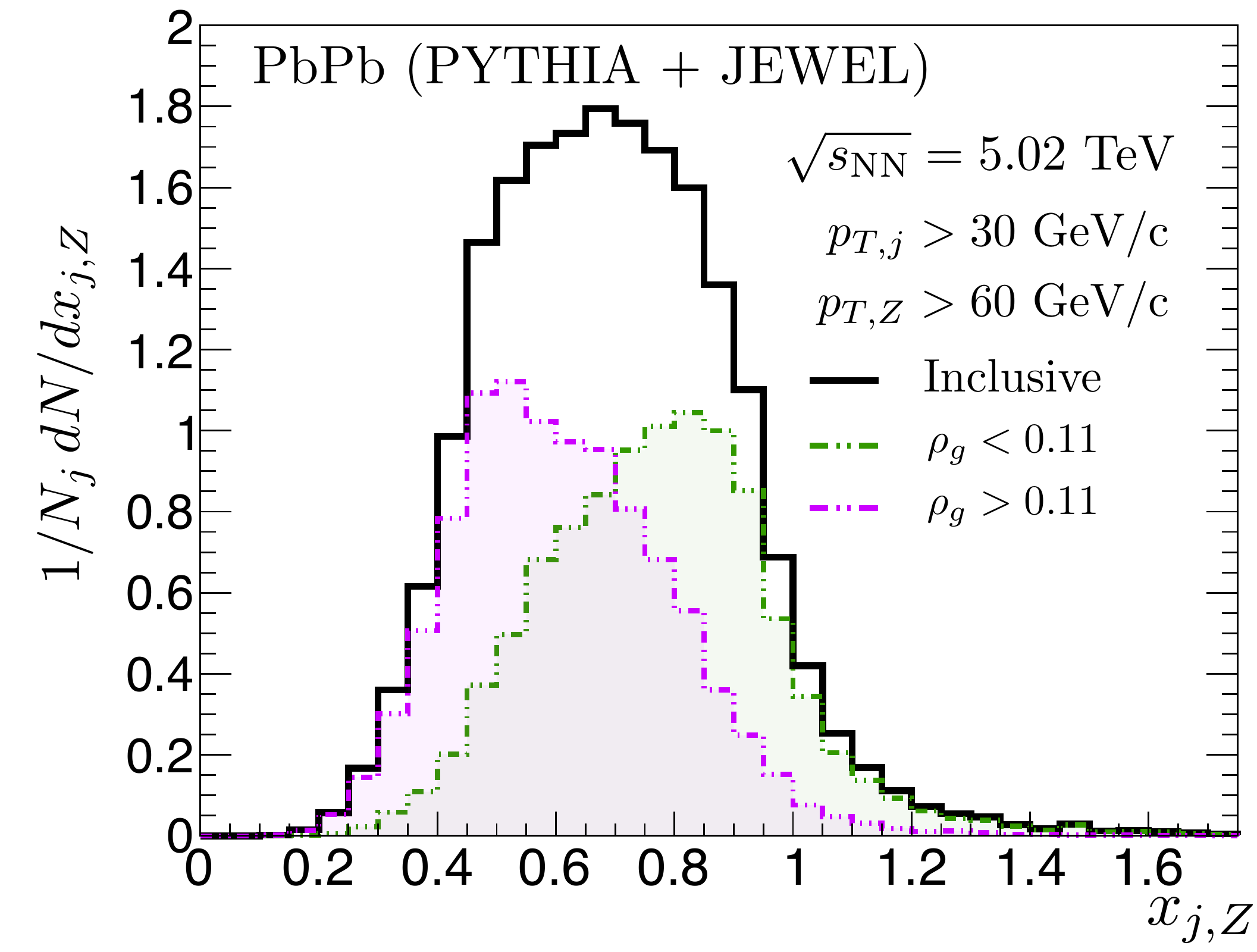}
\caption{Groomed jet mass ($\rho_g$) of the first $\tau$-declustering step that satisfies $z_g>z_\text{cut}$ (top panel) and boson-jet momentum asymmetry (\xjz, bottom panel) distributions obtained from $\sqrt{s_{NN}} = 5.02~\rm{TeV}$ Z+jet JEWEL+PYTHIA PbPb events, normalized to the total number of jets in the sample. The violet distribution refers to the $50\%$ heavier jets, while the green distribution characterizes the $50\%$ lighter jets. The median of $\rho_g$ in this sample yields 0.11.}
\label{fig:mg_bins_PbPb}
\end{figure}

Following the same steps as the previous section, we start by examining the correlation between \tauf~and $\rho_g$, shown in \figref{fig:correl_tau_mg}. Again, we focus on JEWEL+PYTHIA PbPb collisions, as the pp case yields very similar results. As expected, small formation times are strongly correlated to large jet masses, whereas larger values of \tauf~typically correspond to lighter jets. We also notice a somewhat larger dispersion as that shown in \figref{fig:correl_tau_deltaR}. Despite this smearing, the correlation (along with the analytical dependence discussed above) suggests that \tauf~and $\rho_g$ could provide a redundant characterization of jets (in which case one might opt to use $\rho_g$ instead). In order to investigate this possibility, we repeat the exercise performed in sections \ref{sec:tau_alg} and \ref{sec:tau_vs_deltaR}: we split the jet population into two samples of equal size ($50\%$ heavier and $50\%$ lighter jets in this case) and then, we examine their separate contributions to the boson-jet momentum imbalance distribution. The results are shown in Figs.\ \ref{fig:mg_bins_pp} and \ref{fig:mg_bins_PbPb} for pp and PbPb respectively.

As expected, the $\rho_g$-based jet selection produces $x_{j,Z}$ distributions that are qualitatively equivalent to those obtained with \tauf~and \dR, in both the pp and PbPb cases. The shift between the \xjz~distributions for light and heavy jets is, nonetheless, more pronounced already in pp collisions. Also, PbPb jets with a higher $\rho_g$ (purple dot-dashed line) show a larger $x_{j,Z}$ shift with respect to the pp reference than the lighter ones (green dot-dashed line). 

\begin{figure}
\centering
  \includegraphics[width=.9\linewidth]{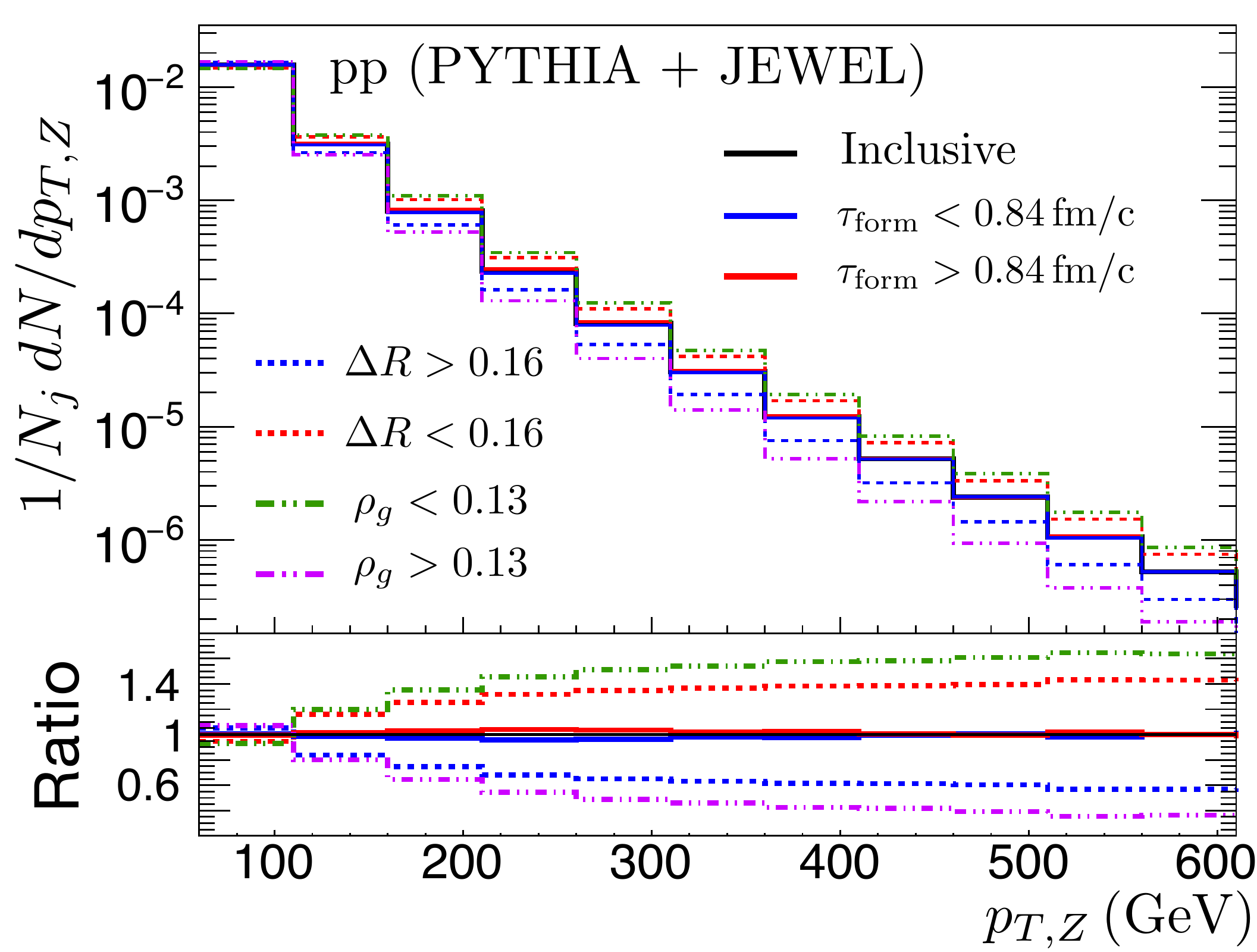}
  \includegraphics[width=.9\linewidth]{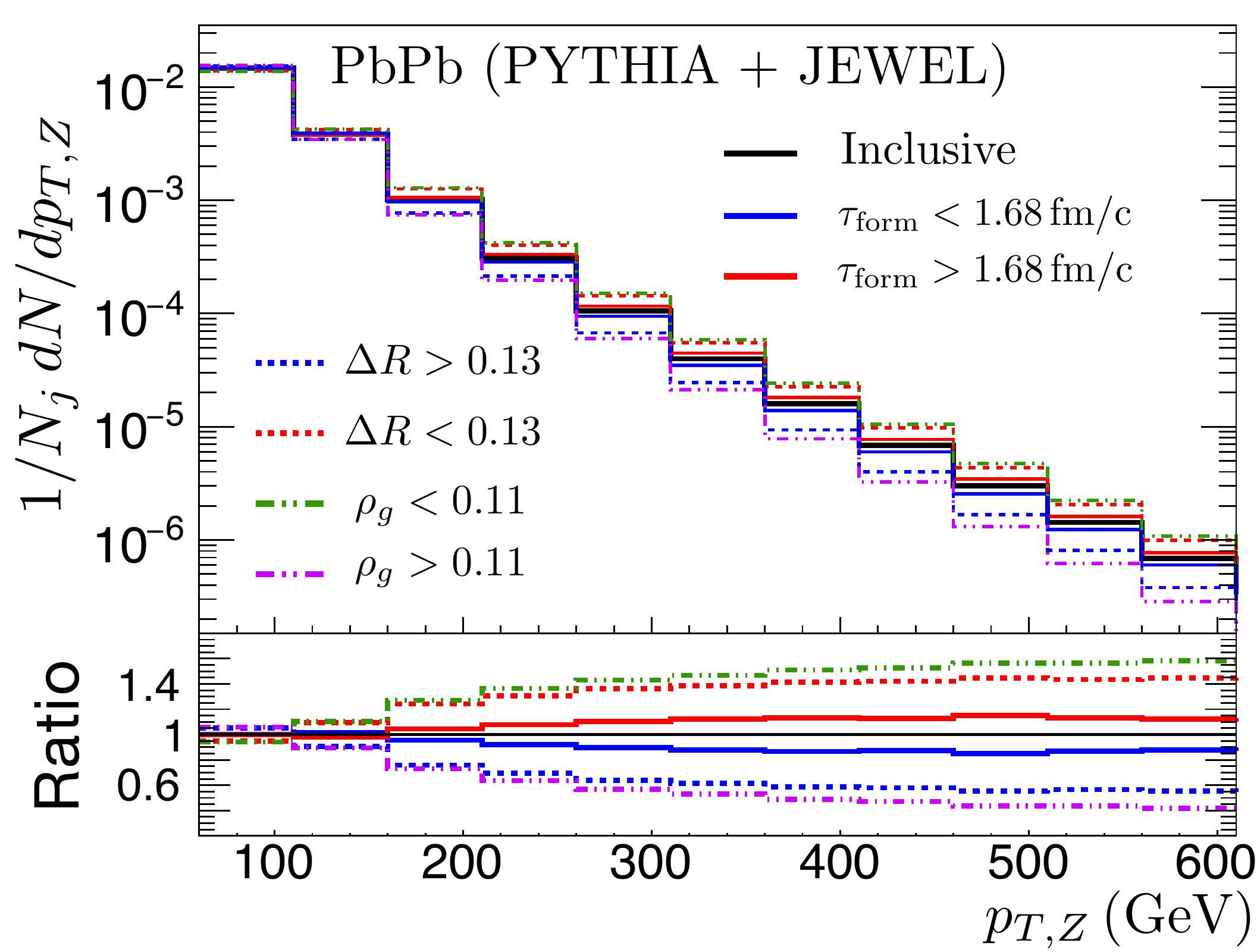}
\caption{Same as Figs.\ \ref{fig:bias_plots_50}, but including the $p_{T,Z}$ spectra for \textit{heavy} and \textit{light} jet samples.}
\label{fig:bias_plots_50_v2}
\end{figure}

As we did in \secref{sec:tau_vs_deltaR}, we explored the differences between these variables by analyzing the $p_{T,Z}$ spectra corresponding to the jet samples they define. The results are shown in \figref{fig:bias_plots_50_v2}, now including the three selectors. By interpreting the $p_{T,Z}$ as a proxy for the initial $p_T$ of the jet, we conclude that the \textit{heavy} (\textit{light}) selection induces a bias towards jets with softer (harder) initial $p_T$. The magnitude of this bias is comparable to that observed in the \dR-based selections.

\section{Conclusions and outlook}
\label{sec:fin}

In this work we apply jet substructure tools to study energy loss in PbPb collisions. We focus on the steps along the jet re-clustered tree as proxies for the splitting along the main branch of the parton showers. This correlation was previously explored in Ref.\ \cite{Apolinario:2020uvt} for the generalized $k_T$ algorithm family, and was found to be optimal when combining the $\tau$ algorithm ($p\!=\!0.5$) with $z_{g} > 0.1$. Using this framework, we extract the formation time \tauf~corresponding to the first emission of the parton shower, and use this quantity as a classifier.

We focus on Z-tagged jets generated using JEWEL v2.2.0 with no recoils, as they allow for a straightforward quantification of energy loss as a momentum imbalance between jet and boson. The resulting distributions display the expected behavior, with the early-splitting jets losing more energy on average than those that start fragmenting later. Each jet sample was defined to contain approximately half of the entire population, with the difference between `early' and `late' being determined by the median of the corresponding \tauf~distribution.

Although we focus our study on \tauf, we notice that similar results may be obtained by labelling jets with the opening angle, \dR, or the (normalized) groomed jet mass, $\rho_g$. A distinct feature between the three variables as jet classifiers lies on the reconstructed $p_{T,Z}$ distribution, used as proxy for the initial hard kinematics originating the jet. Particularly, we observe that the $\Delta R$-based jet selection induces a significant bias on the initial $p_{T}$ distribution of the jets, both in pp and PbPb collisions, with the `wide' (`narrow') jet sample being biased towards softer (harder) initial $p_T$. Qualitatively similar results are found for the $\rho_g$-based samples, with the `heavy' (`light') sample featuring a softer (harder) initial $p_T$ spectra. Such biases complicate the interpretation of energy loss observables, as differences in the reconstructed final jet are a convolution of energy loss effects and a change in the initial hard kinematics. On the other hand, the initial $p_{T}$ distributions of the $\tau_{form}$-based samples are remarkably closer to the inclusive spectrum (being unaffected in pp collisions). This allows for a more meaningful comparison between two distinct jet classes, as their differences are dominated by (different levels of) quenching effects rather than kinematic biases. Finally, we note that this exercise constitutes a further check as to whether $\tau_{form}$~(as defined through the $\tau$ algorithm) can be effectively identified as an internal scale of the parton shower independently of the initial momenta of jets. We intend to explore this further in a subsequent publication.

We considered here $Z$+jet events with the aim of quantitatively understanding kinematic biases in the jet-initiating parton distribution introduced by selecting different jet populations. We expect that our finding - classifying jets according to \tauf\ significantly reduces such biases - directly carries over to dijet events. Studies comparing jet populations with different formation times can therefore be carried out on the much larger single-inclusive jet event samples.

Although the proposed method would in principle allow access to all emissions along the primary branch, in this paper we restrict ourselves to using the first unclustering step only. The feasibility of extracting meaningful physical information from subsequent splitting times requires careful consideration and study, which we leave for future work.

\vspace{2mm}
\textbf{Acknowledgments}
    The authors would like to thank H. Bossi, R. Ehlers and L. Havener for fruitful discussions. LA and PGR also would like to thank J. Nesbitt for his valuable contribution to a parallel but related project during the LIP Summer Internship program. This work was supported by European Research Council (ERC) projects ERC-2018-ADG-835105 YoctoLHC and ERC-2018-STG-803183 CollectiveQCD; by OE Portugal, Funda\c{c}\~{a}o para a Ci\^{e}ncia e a Tecnologia (FCT), I.P., projects EXPL/FIS-PAR/0905/2021 and CERN/FIS-PAR/0032/2021. L.A. acknowledges the financial support by FCT under contract 2021.03209.CEECIND.

\appendix

\section{\eqref{eq:tauf} vs \eqref{eq:tauf2}}\label{app:tau_comp}

\begin{figure}
\centering
  \includegraphics[width=.9\linewidth]{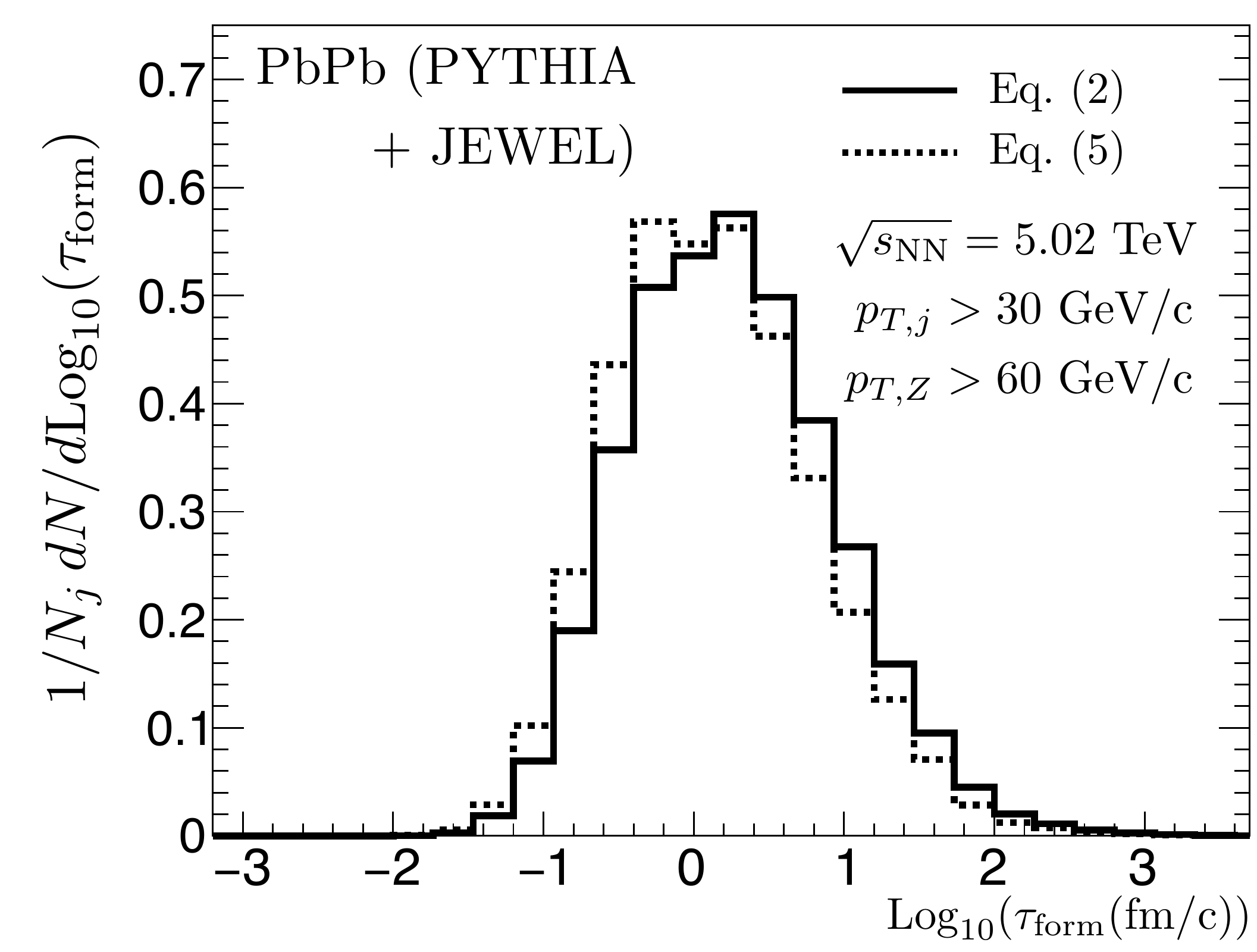}
\caption{Comparison of two alternative estimations of the formation time (\tauf) of the first $\tau$-declustering step that satisfies $z_g>z_\text{cut}$, obtained from $\sqrt{s_{NN}} = 5.02~\rm{TeV}$ Z+jet JEWEL+PYTHIA PbPb events.}
\label{fig:tau_comparison}
\end{figure}

For completeness, in this appendix, we illustrate the differences on the extracted $\tau_{form}$ when using \eqref{eq:tauf} or, alternatively, when the collinear limit is used instead to write $\tau_{form}$ as in \eqref{eq:tauf2}. Both are calculated with the kinematics of the first two subjets to survive the $z_g > z_{cut}$ condition on a $\tau$-re-clustered tree. The results for PbPb events are shown in \figref{fig:tau_comparison} with Eqs.~(\ref{eq:tauf}) and~(\ref{eq:tauf2}) represented in solid and dashed lines respectively. Although not explicitly included in this appendix, pp events also display the same differences. 

When employing a collinear approximation, the obtained $\tau_{form}$ will naturally be shifted towards smaller values, although the differences are not significant. This allows to, within this precision, to use interchangeably the two $\tau_{form}$ definitions.

\section{$\tau$- vs C/A-re-clustered jets}
\label{app:ca_vs_tau}

\begin{figure}
\centering
  \includegraphics[width=.9\linewidth]{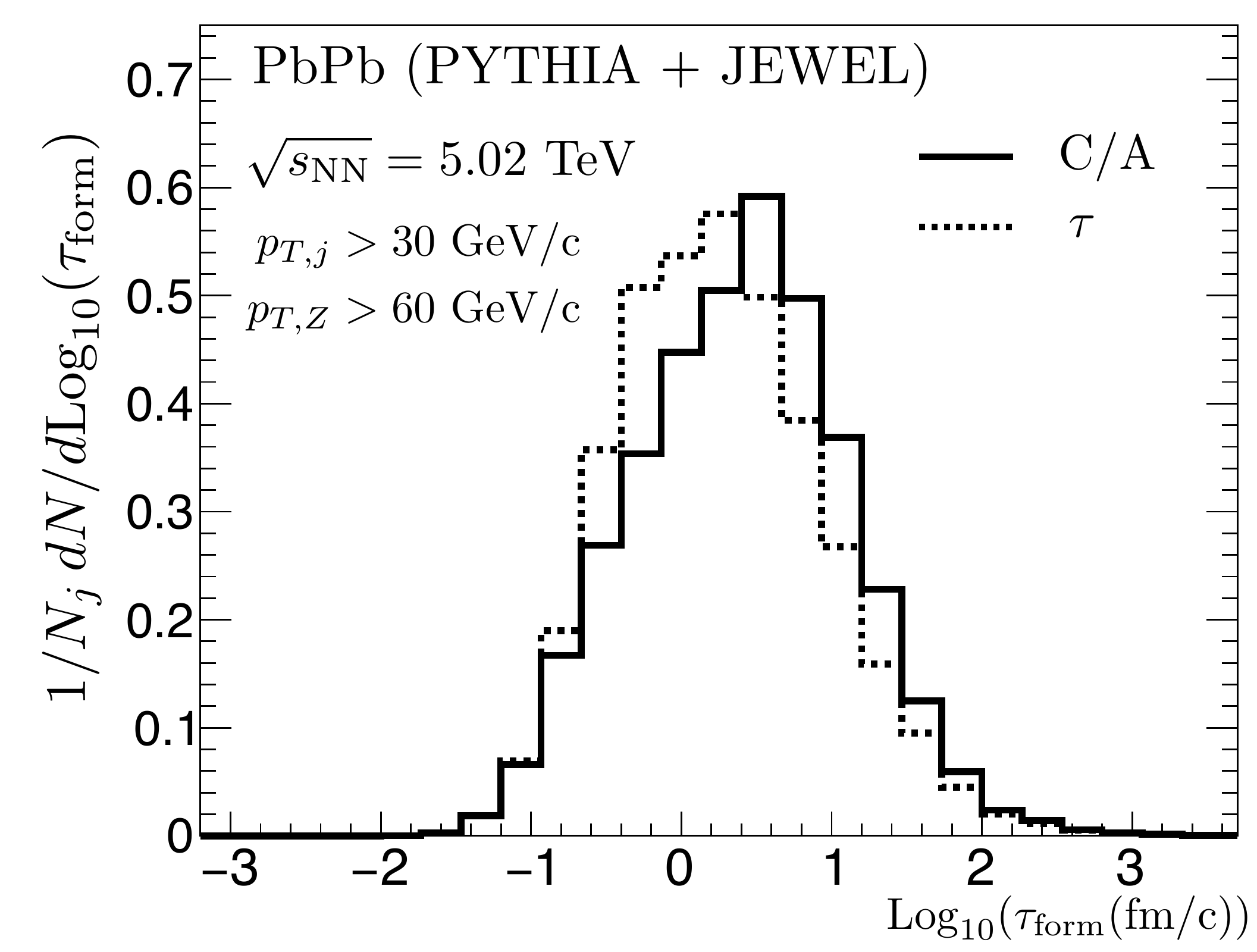}
  \includegraphics[width=.9\linewidth]{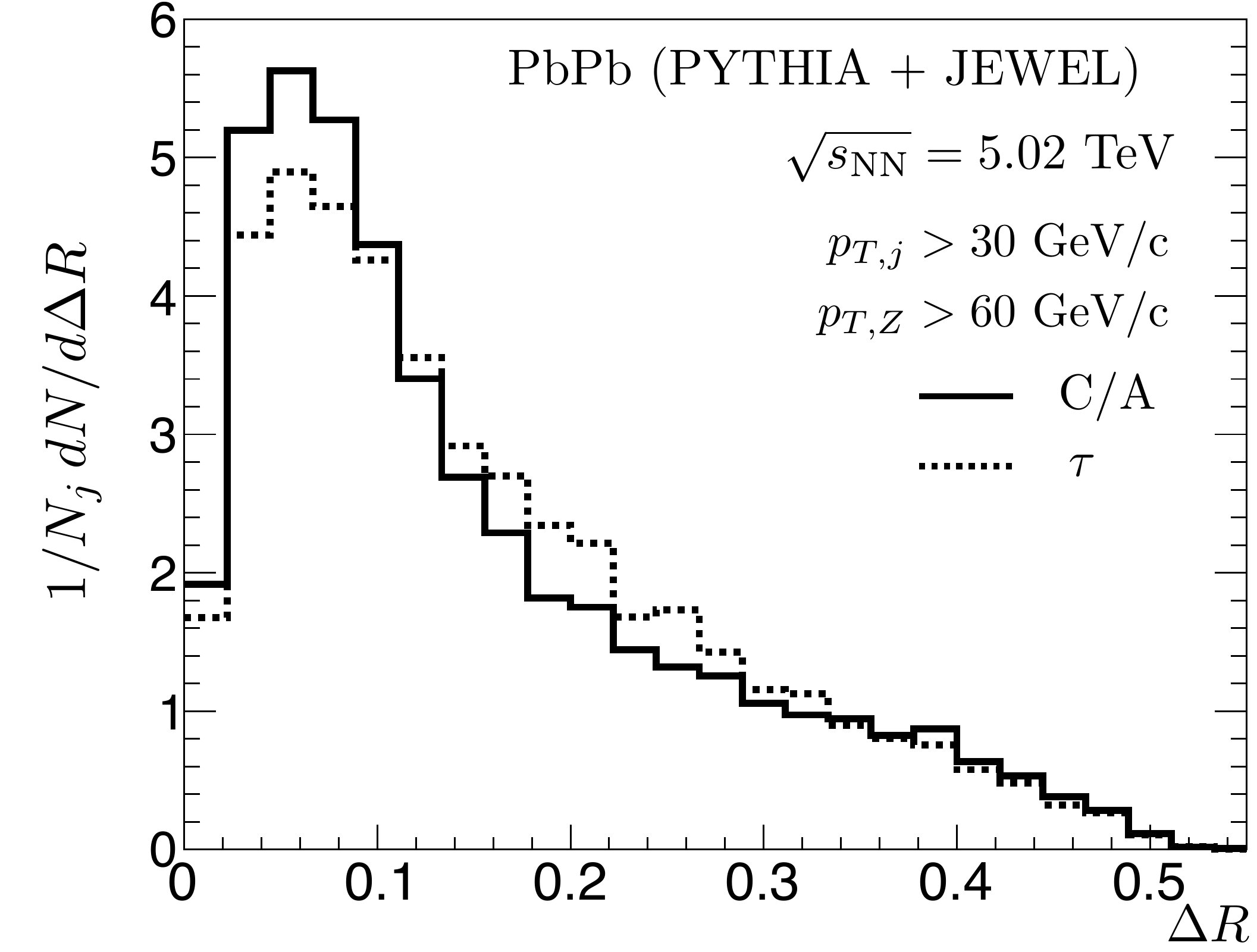}  \caption{Distributions for \tauf~(top panel) and \dR~(bottom panel) of the clustering step to survive $z_g > z_{cut}$ on a jet re-clustered tree obtained via C/A (solid line) and $\tau$ (loosely dashed line) algorithms. The jets were obtained from $\sqrt{s_{NN}} = 5.02~\rm{TeV}$ Z+jet JEWEL+PYTHIA PbPb events.}
\label{fig:algorithm_comparison1_a}
\end{figure}

In \figref{fig:algorithm_comparison1_a} we show a comparison of the \tauf~and \dR~distributions extracted by re-clustering PbPb jets with C/A (solid lines) and $\tau$ (loosely dashed lines) algorithms. For brevity's sake, in this appendix we focus on PbPb events only, as the same discussion applies to the pp case.

As expected, the C/A algorithm provides slightly larger values of \tauf~on average, as it typically favours splittings characterized by lower values of $z$ (see \eqref{eq:tauf}) and larger \dR~values. Despite these differences, we observe that the analysis ensuing from separating these distributions by their medians yields identical conclusions in both cases. The reason for this is that, albeit somewhat different in shape, the amount of jets that swap classes when considering one or other algorithm is negligible.

\begin{figure}
\centering
   \includegraphics[width=.9\linewidth]{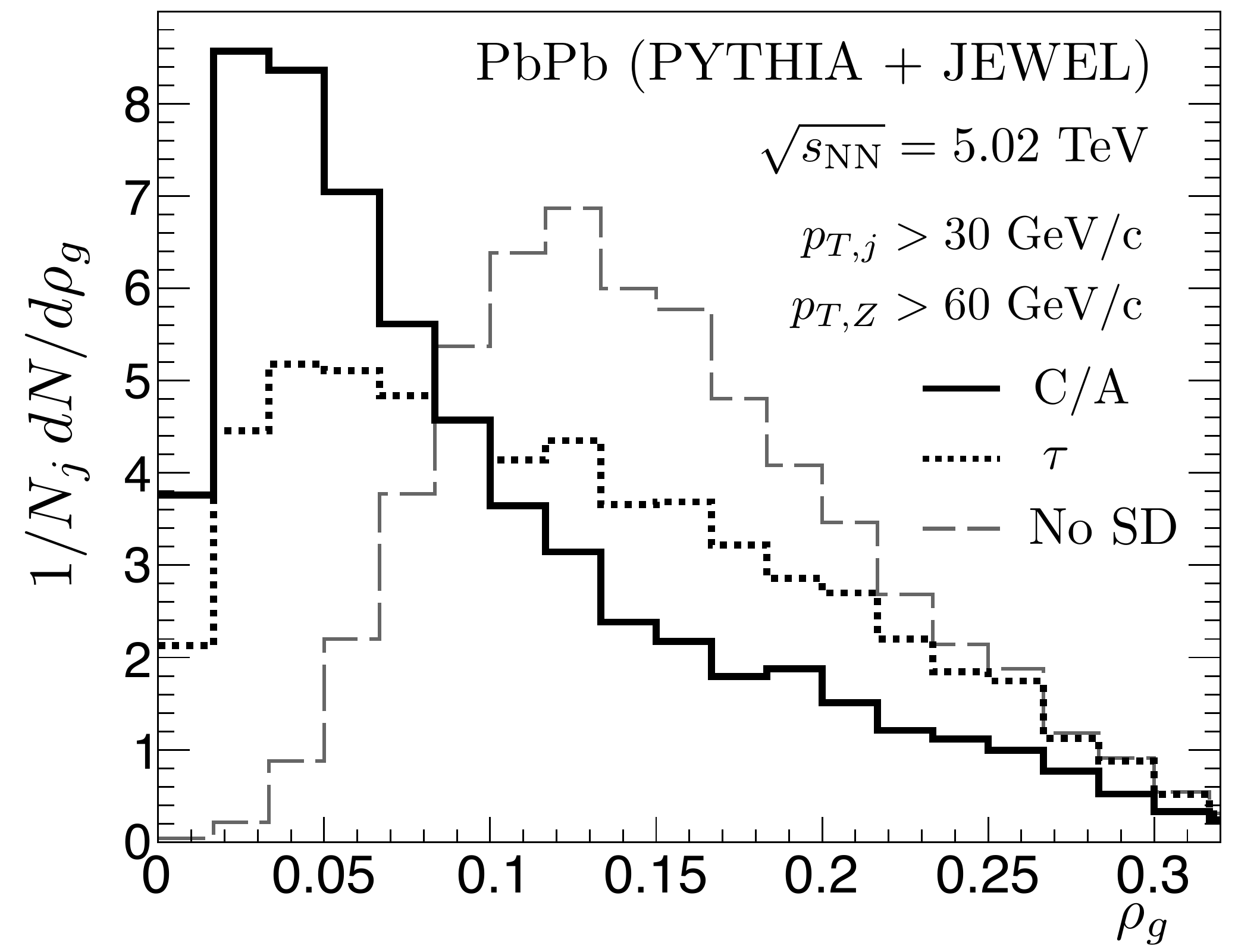}
  \caption{Distributions for $\rho_g$ of the clustering step to survive $z_g > z_{cut}$ on a jet re-clustered tree obtained via C/A (solid line) and $\tau$ (loosely dashed line) algorithms. In addition, we also include the un-groomed $\rho_g$ distribution (densely dashed line). The jets were obtained from $\sqrt{s_{NN}} = 5.02~\rm{TeV}$ Z+jet JEWEL+PYTHIA PbPb events.}
\label{fig:algorithm_comparison1_b}
\end{figure}

A different situation is observed for the $\rho_g$ case, illustrated in \figref{fig:algorithm_comparison1_b} along with the un-groomed mass distribution, for reference (densely dashed line). For this variable, both the median and overall shape of the distribution bear a significant dependence on the algorithm employed in the grooming process (see appendix~\ref{app:grooming} for more details). However, we observe that this sensitivity is not reflected in our final results, as our selection is solely based on the quantiles of the distributions. This allowed us to focus exclusively on the $\tau$ algorithm in the main body of the paper.

\section{Effect of grooming}
\label{app:grooming}

\begin{figure}
\centering
  \includegraphics[width=0.95\linewidth]{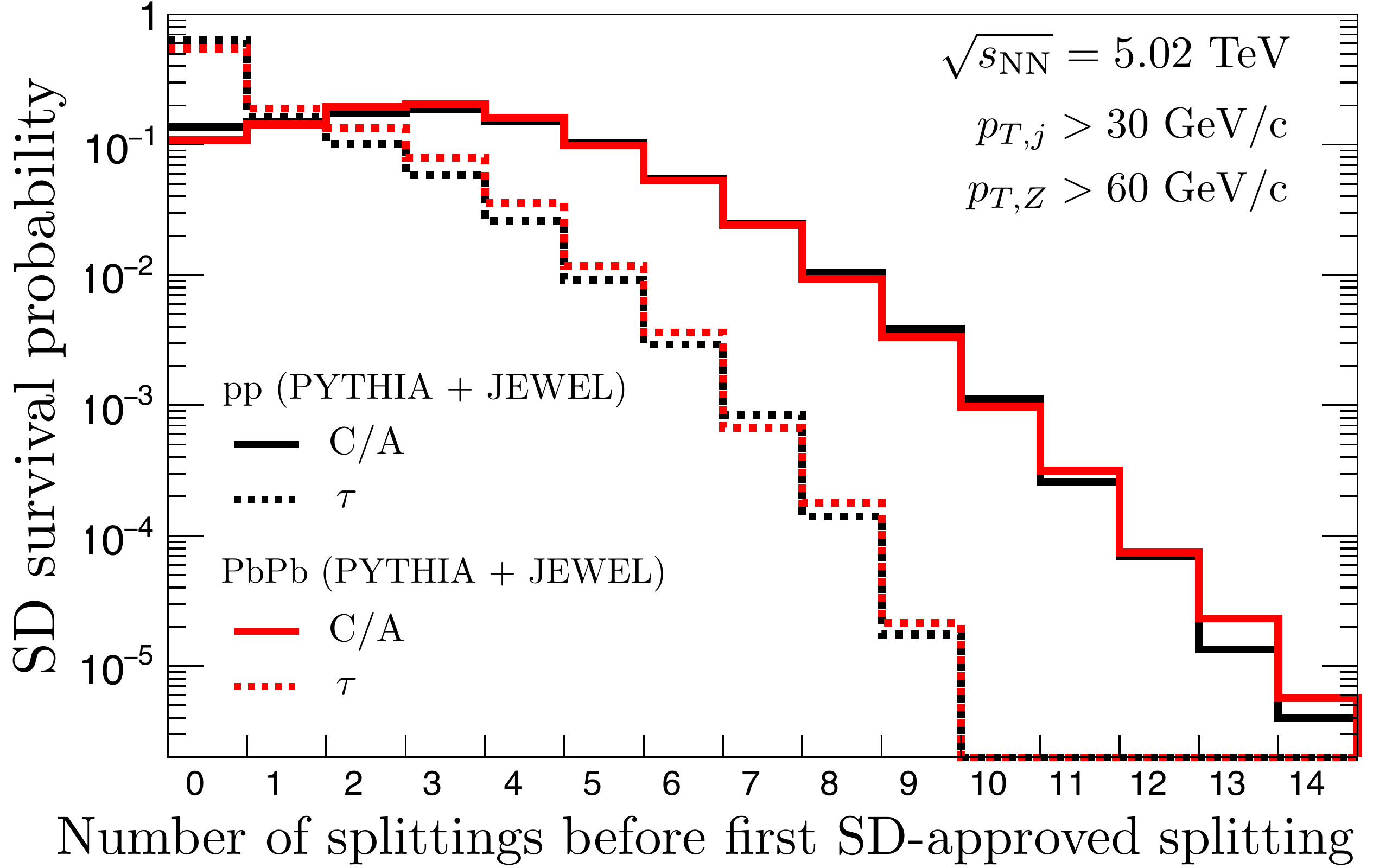}
\caption{Probability of an event containing at least one splitting that survives the condition $z_g > z_{cut}$ as a function of the amount of previously discarded splittings. Solid (dashed) curves correspond to events re-clustered with C/A ($\tau$) algorithm. Black (blue) curves correspond to pp (PbPb) events.}
\label{fig:SDprob}
\end{figure}

In \figref{fig:SDprob} we show the probability of finding an event with $z_g>z_\text{cut}$ (SD survival probability), plotted as a function of the amount of splittings discarded before this criteria is met.

The C/A re-clustering is represented in solid lines, while the $\tau$ re-clustered jets are depicted as dashed lines. The pp and PbPb distributions are shown in black and red, respectively. As expected, the number of clustering steps discarded in C/A until one meets SD criteria is considerably larger, as these typically represent unphysical clusterings of single particles near the boundaries of the jet area. As such, to make the comparison between the jet mass and the remaining selection variables (\tauf~and \dR), we chose to always impose the SD condition, regardless of the re-clustering procedure. As a consequence, the jet mass distribution, which is known to be very sensitive to non-perturbative effects, is drastically changed by the chosen grooming settings. In \figref{fig:algorithm_comparison1_b} we show the groomed mass $\rho_g$ distribution for $\tau$- (dashed line) and C/A- (solid line) re-clustered jets when $z_g>z_\text{cut}$ is imposed. For reference, as a densely dashed line, it is illustrated the ungroomed jet mass.

\figref{fig:algorithm_comparison1_b} suggests that significantly more mass is groomed away when applying C/A re-clustering, which leads to a lighter, more sharply peaked $\rho_g$ distribution. We attribute this discrepancy to the enhanced sensitivity of the C/A algorithm to soft contamination. When using the $\tau$ algorithm, identifying a splitting with $z_g>z_\text{cut}$ within the first few emissions is quite likely. However, the probability of this happening drops substantially for the purely geometry-based C/A algorithm, which typically discards several large-angle emissions first (see \figref{fig:SDprob}). Nonetheless, we reiterate that, despite the different shapes, the conclusions based on the quantiles of the distributions are kept unchanged. 

\section{Smaller jet selection quartiles}
\label{app:bias_smaller_selections}

In this appendix we experiment with more restrictive jet selection criteria. Specifically, we re-define the early/late jet classification to be based upon the $Q_1$ and $Q_3$ quartiles of the \tauf~distribution rather than its median. The same method goes for the wide/narrow and heavy/light samples. In doing so, we aim to test whether the biases increase with a smaller jet selection. 

\begin{figure} [b]
\centering
\includegraphics[width=0.9\linewidth]{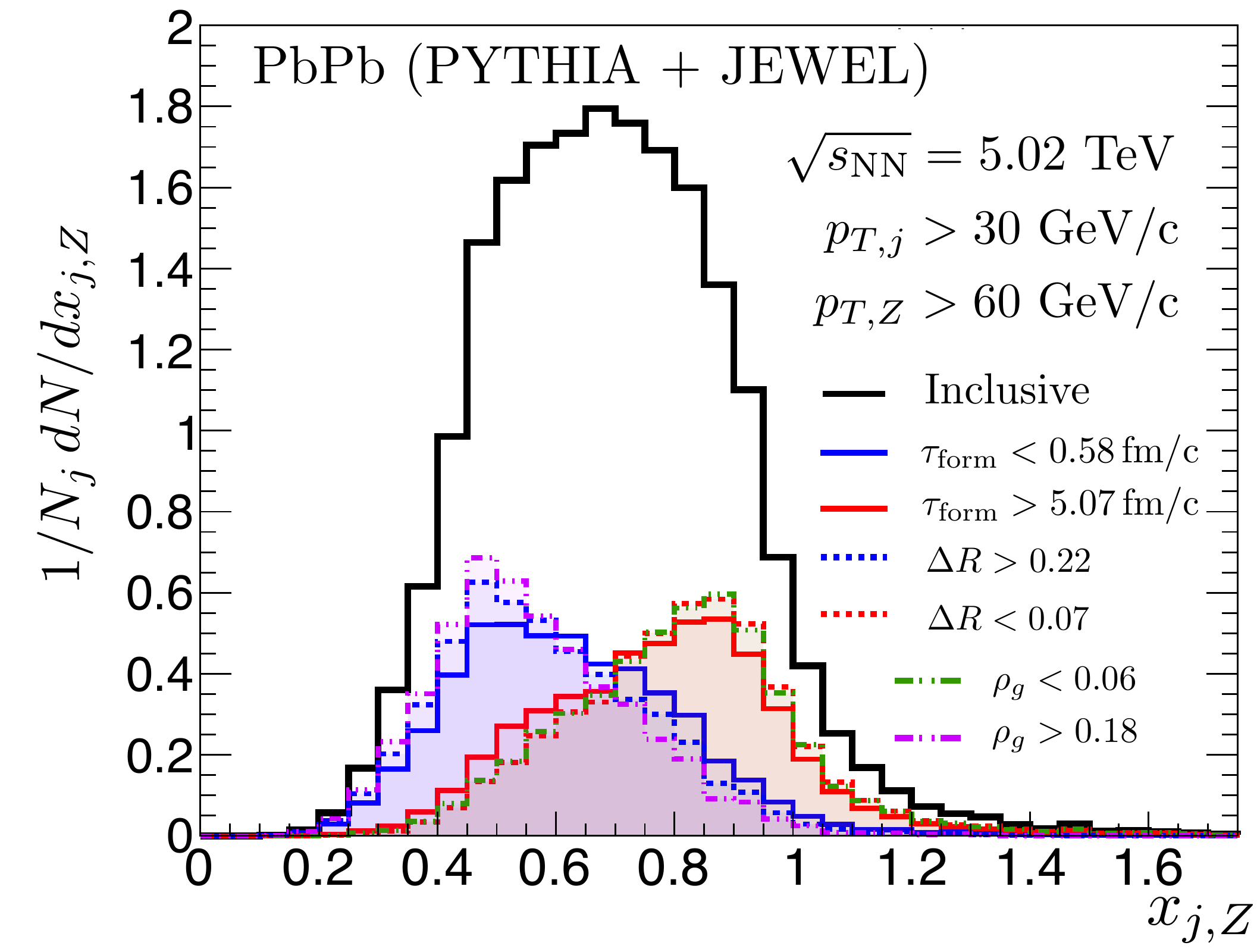}
\caption{Same as bottom panel of \figref{fig:dR_bins_PbPb} and \figref{fig:mg_bins_PbPb}, but with each sample containing $25\%$ of the total jet population.}
\label{fig:bias_plots_20}
\end{figure}

\begin{figure*}
\centering
\includegraphics[width=0.43\linewidth]{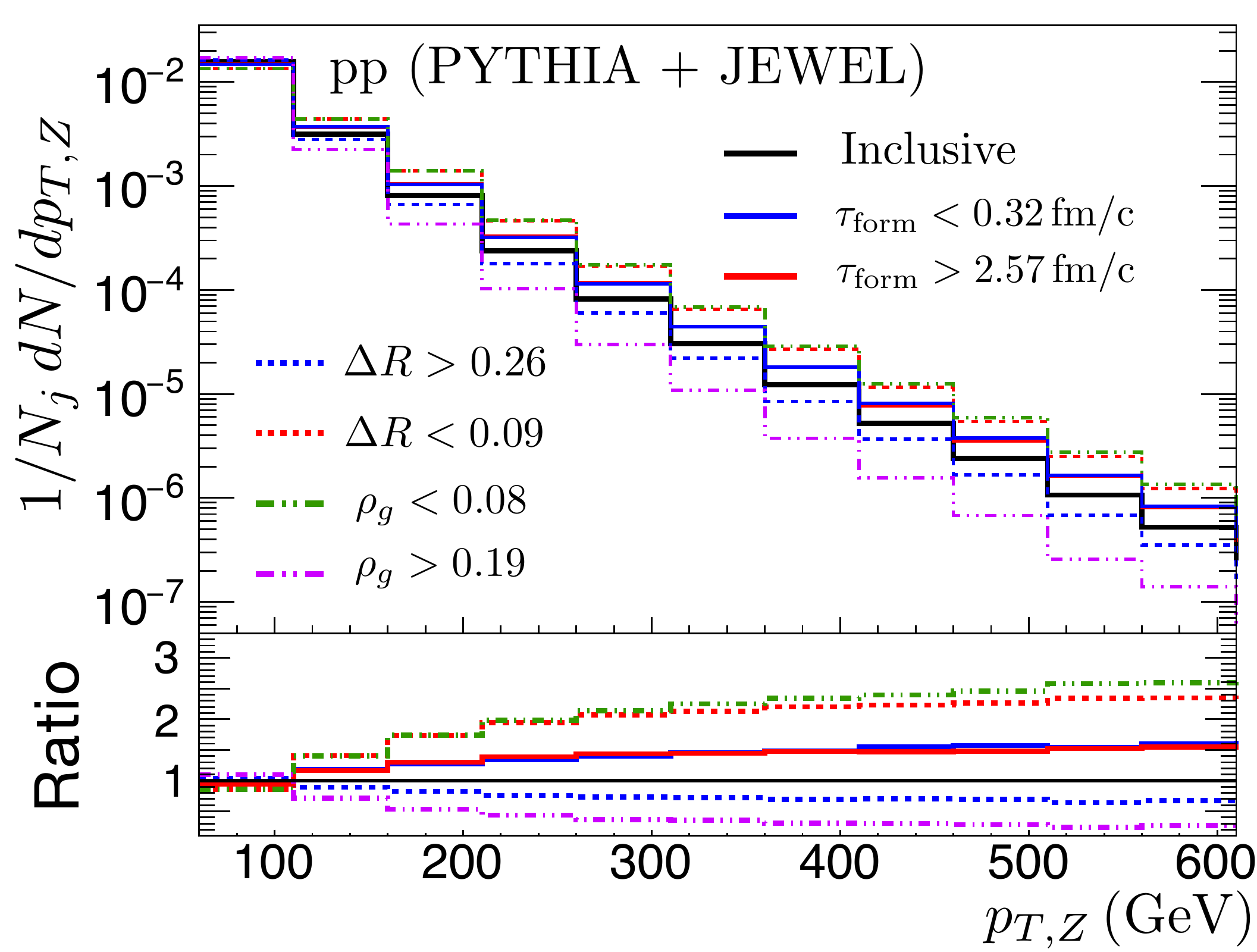}
\includegraphics[width=0.43\linewidth]{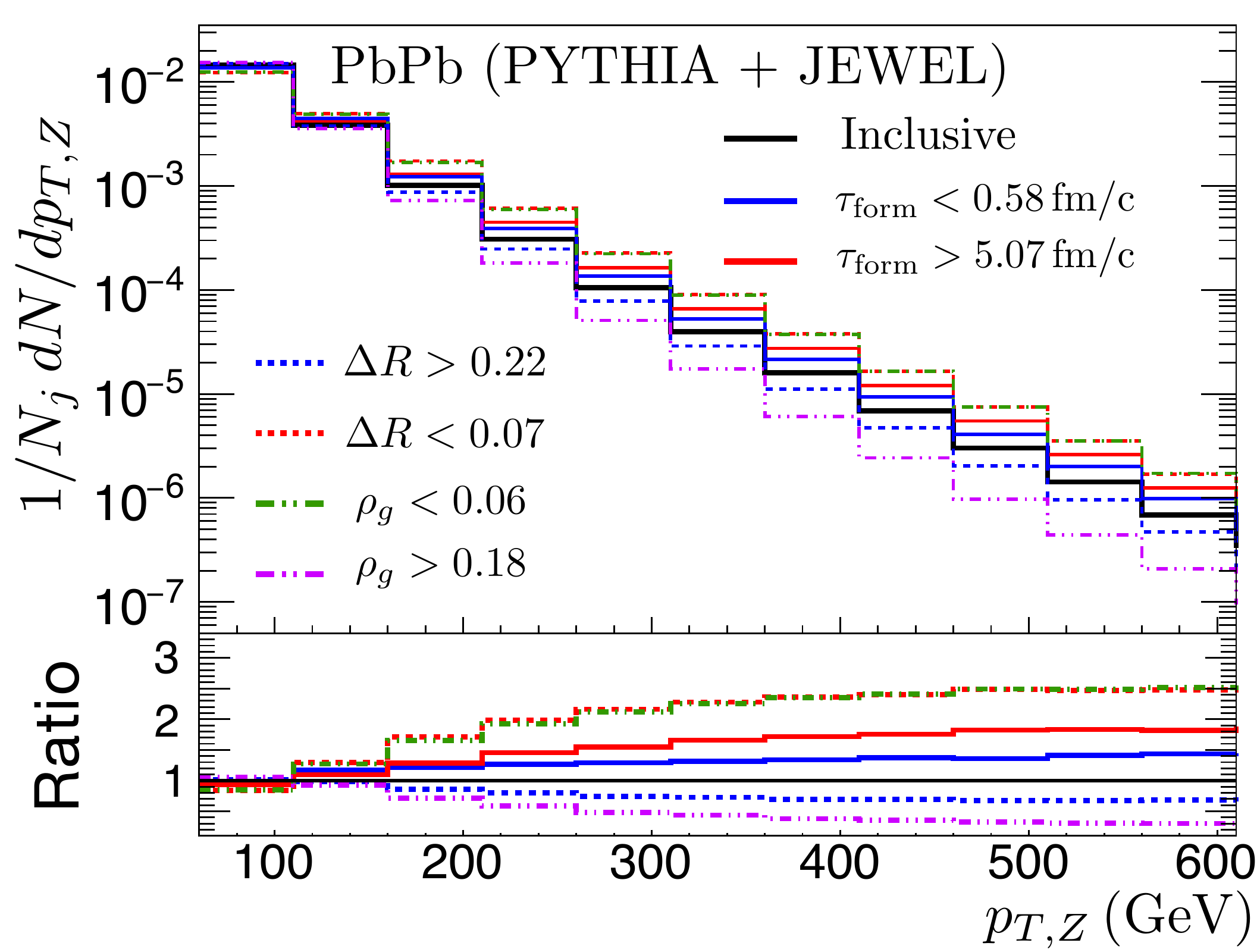}
\caption{Same as \figref{fig:bias_plots_50_v2}, but with each sample containing $25\%$ of the total jet population.}
\label{fig:bias_plots_25}
\end{figure*}

The equivalent of the bottom panel of \figref{fig:dR_bins_PbPb} is now shown in \figref{fig:bias_plots_20}, but considering the $25\%$ selections for the early/late (in solid), wide/narrow (in dashed) and light/heavy jets (in dot-dashed). Doing this selection in the pp reference does not yield any significant shift between the two populations as previously found when considering a classification based on the median, $Q_2$ (bottom panels of \figref{fig:dR_bins_pp} and \figref{fig:mg_bins_pp}). For PbPb jets however, it continues to have a clear separation between the two samples. In fact, the most noticeable effect is simply the decrease in the size of the samples. Nonetheless, making a smaller selection on the jet population does strongly impact over the corresponding $p_{T,Z}$ spectra for both pp and PbPb events, as shown in \figref{fig:bias_plots_25} left and right panel respectively. In this plot we observe that, by performing a more restrictive jet selection, the resulting populations become generally more biased with respect to the inclusive scenario. However, we do note that \tauf~still provides the smaller bias when accounting for the two selected intervals. 

Interestingly, the early jet selection within QGP falls within the pre-QGP phase ($\tau_{\text{form}} \lesssim 0.5~\rm{fm/c}$), where a consistent understanding of in-medium interactions is still lacking. Future studies focused on this selection using the $\tau$ algorithm together with a selection based on $\tau_{\text{form}}$ can potentially be used to provide further insights into this stage of the collision evolution. This and other similar studies will be pursued in a future communication. 

\bibliographystyle{JHEP-2modlong}
\bibliography{main}

\end{document}